\documentclass{aastex631}

\usepackage{graphicx,multirow,amssymb}

\accepted{January 13, 2025}

\begin{document}

\title{Stellar occultation observations of (38628) Huya and its satellite: a detailed look into the system}

\author[0000-0002-6085-3182]{F. L. Rommel}
\affiliation{Florida Space Institute, 12354 Research Parkway, Partnership I, Room 211, 32826 Orlando, United States of America}
\affiliation{Federal University of Technology - Paraná (PPGFA/UTFPR-Curitiba), Av. Sete de Setembro 3165, 80230-901 Curitiba, Brazil}
\affiliation{Laboratório Interinstitucional de e-Astronomia (LIneA/INCT do e-Universo), Av. Pastor Martin Luther King Jr, 126, 20765, Rio de Janeiro, Brazil}
\correspondingauthor{Flavia L. Rommel}
\email{flavialuane.rommel@ucf.edu}

\author[0000-0003-2132-7769]{E. Fernández-Valenzuela}
\affiliation{Florida Space Institute, 12354 Research Parkway, Partnership I, Room 211, 32826 Orlando, United States of America}

\author[0000-0002-1788-870X]{B. C. N. Proudfoot}
\affiliation{Florida Space Institute, 12354 Research Parkway, Partnership I, Room 211, 32826 Orlando, United States of America}

\author[0000-0002-8690-2413]{J. L. Ortiz} 
\affiliation{Instituto de Astrofísica de Andalucía (CSIC), Glorieta de la Astronomía s/n, 18008 Granada , Spain}

\author[0000-0003-0088-1808]{B. E. Morgado}
\affiliation{Observatório do Valongo (OV/UFRJ), Ladeira do Pedro Antônio 43, 20080 Rio de Janeiro, Brazil}

\author[0000-0003-1995-0842]{B. Sicardy}
\affiliation{IMCCE, Observatoire de Paris, PSL Université, Sorbonne Université, Université Lille 1, CNRS UMR 8028, 77 avenue Denfert-Rochereau, 75014 Paris, France}

\author[0000-0003-0419-1599]{N. Morales} 
\affiliation{Instituto de Astrofísica de Andalucía (CSIC), Glorieta de la Astronomía s/n, 18008 Granada , Spain}

\author[0000-0003-2311-2438]{F. Braga-Ribas}
\affiliation{Federal University of Technology - Paraná (PPGFA/UTFPR-Curitiba), Av. Sete de Setembro 3165, 80230-901 Curitiba, Brazil}
\affiliation{Laboratório Interinstitucional de e-Astronomia (LIneA/INCT do e-Universo), Av. Pastor Martin Luther King Jr, 126, 20765, Rio de Janeiro, Brazil}

\author[0000-0002-2193-8204]{J. Desmars}
\affiliation{IMCCE, Observatoire de Paris, PSL Université, Sorbonne Université, Université Lille 1, CNRS UMR 8028, 77 avenue Denfert-Rochereau, 75014 Paris, France}

\author[0000-0003-1690-5704]{R. Vieira-Martins}
\affiliation{Observatório Nacional (ON/MCTI), R. General José Cristino 77, 20921 Rio de Janeiro, Brazil}

\author[0000-0002-6117-0164]{B. J. Holler}
\affiliation{Space Telescope Science Institute, Steven Muller Building, 3700 San Martin Drive, 21218 Baltimore, United States of America}

\author[0000-0001-8641-0796]{Y. Kilic}
\affiliation{Instituto de Astrofísica de Andalucía (CSIC), Glorieta de la Astronomía s/n, 18008 Granada , Spain}

\author[0000-0002-8296-6540]{W. Grundy}
\affiliation{Lowell Observatory,86001 Flagstaff, Arizona, United States of America}
\affiliation{Northern Arizona University, 86001 Flagstaff, Arizona, United States of America}

\author[0000-0002-9789-1203]{J. L. Rizos}
\affiliation{Instituto de Astrofísica de Andalucía (CSIC), Glorieta de la Astronomía s/n, 18008 Granada , Spain}

\author[0000-0002-1642-4065]{J. I. B. Camargo}
\affiliation{Observatório Nacional (ON/MCTI), R. General José Cristino 77, 20921 Rio de Janeiro, Brazil}
\affiliation{Laboratório Interinstitucional de e-Astronomia (LIneA/INCT do e-Universo), Av. Pastor Martin Luther King Jr, 126, 20765, Rio de Janeiro, Brazil}

\author[0000-0002-4106-476X]{G. Benedetti-Rossi}
\affiliation{Laboratório Interinstitucional de e-Astronomia (LIneA/INCT do e-Universo), Av. Pastor Martin Luther King Jr, 126, 20765, Rio de Janeiro, Brazil}

\author[0000-0002-3362-2127]{A. Gomes-Júnior}
\affiliation{Universidade Federal de Uberlândia (UFU), 38408 Uberlândia, Brazil}

\author[0000-0002-8211-0777]{M. Assafin}
\affiliation{Observatório do Valongo (OV/UFRJ), Ladeira do Pedro Antônio 43, 20080 Rio de Janeiro, Brazil}
\affiliation{Laboratório Interinstitucional de e-Astronomia (LIneA/INCT do e-Universo), Av. Pastor Martin Luther King Jr, 126, 20765, Rio de Janeiro, Brazil}

\author[0000-0002-1123-983X]{P. Santos-Sanz} 
\affiliation{Instituto de Astrofísica de Andalucía (CSIC), Glorieta de la Astronomía s/n, 18008 Granada , Spain}

\author[0000-0001-8858-3420]{M. Kretlow} 
\affiliation{Instituto de Astrofísica de Andalucía (CSIC), Glorieta de la Astronomía s/n, 18008 Granada , Spain}

\author[0000-0002-8112-0770]{M. Vara-Lubiano} 
\affiliation{Instituto de Astrofísica de Andalucía (CSIC), Glorieta de la Astronomía s/n, 18008 Granada , Spain}

\author[0000-0002-6477-1360]{R. Leiva} 
\affiliation{Instituto de Astrofísica de Andalucía (CSIC), Glorieta de la Astronomía s/n, 18008 Granada , Spain}

\author[0000-0003-1080-9770]{D. A. Ragozzine}
\affiliation{Department of Physics and Astronomy, N283 ESC, Brigham Young University, 84602 Provo, Utah, United States of America}

\author[0000-0001-5963-5850]{R. Duffard}
\affiliation{Instituto de Astrofísica de Andalucía (CSIC), Glorieta de la Astronomía s/n, 18008 Granada , Spain}

\author[0000-0002-1330-1318]{H. Ku\v{c}áková}
\affiliation{Research Centre for Theoretical Physics and Astrophysics, Institute of Physics, Silesian University in Opava, Bezručovo nám. 13, CZ-746 01 Opava, Czech Republic}
\affiliation{Astronomical Institute, Faculty of Mathematics and Physics, Charles University Prague, Praha 8, CZ-180 00 V Holešovičkách 2, Czech Republic}
\affiliation{Astronomical Institute of the Czech Academy of Sciences, Fričova 298, CZ-251 65 Ondřejov, Czech Republic}

\author[0000-0002-0835-225X]{K. Hornoch}
\affiliation{Astronomical Institute of the Czech Academy of Sciences, Fričova 298, CZ-251 65 Ondřejov, Czech Republic}

\author[0009-0007-7900-4811]{V. Nikitin}
\affiliation{International Occultation Timing Association (IOTA), United States of America}

\author[0000-0002-0143-9440]{T. Santana-Ros}
\affiliation{Departamento de Fisica, Ingeniería de Sistemas y Teoría de la Señal, Universidad de Alicante, Carr. de San Vicente del Raspeig, s/n, 03690 San Vicente del Raspeig, Alicante, Spain}
\affiliation{Institut de Ci\`{e}ncies del Cosmos (ICCUB), Universitat de Barcelona (IEEC-UB),
Carrer de Mart\'{\i} i Franqu\`{e}s, 1, 08028 Barcelona, Spain}

\author{O. Canales-Moreno}
\affiliation{Red Astronavarra Sarea, Astrosedetania, Spain}

\author{D. Lafuente-Aznar}
\affiliation{Red Astronavarra Sarea, Astrosedetania, Spain}

\author{S. Calavia-Belloc}
\affiliation{Red Astronavarra Sarea, Astrosedetania, Spain}

\author[0000-0001-9707-2091]{C. Perelló}
\affiliation{Agrupación Astronómica de Sabadell, Carrer Prat de la Riba, 116, 08206 Sabadell, Spain}
\affiliation{International Occultation Timing Association/European Section, Am Brombeerhag 13, 30459 Hannover, Germany}

\author{A. Selva}
\affiliation{Agrupación Astronómica de Sabadell, Carrer Prat de la Riba, 116, 08206 Sabadell, Spain}
\affiliation{International Occultation Timing Association/European Section, Am Brombeerhag 13, 30459 Hannover, Germany}

\author{F. Organero}
\affiliation{Complejo Astronómico de La Hita, Camino de Doña Sol, s/n, 45850 LaVilla de Don fadrique, Spain}

\author{L. A. Hernandez}
\affiliation{Complejo Astronómico de La Hita, Camino de Doña Sol, s/n, 45850 LaVilla de Don fadrique, Spain}

\author{I. de la Cueva}
\affiliation{Astronomical Association of Eivissa (AAE), Lucio Oculacio 29, 07800 Ibiza, Spain}

\author{M. Yuste-Moreno}
\affiliation{Astronomical Association of Eivissa (AAE), Lucio Oculacio 29, 07800 Ibiza, Spain}

\author{E. García-Navarro}
\affiliation{Agrupación Astronómica de Cuenca, Astrocuenca, Spain}

\author{J. E. Donate-Lucas}
\affiliation{Agrupación Astronómica de Cuenca, Astrocuenca, Spain}

\author{L. Izquierdo-Carrión}
\affiliation{Agrupación Astronómica de Cuenca, Astrocuenca, Spain}

\author{R. Iglesias-Marzoa}
\affiliation{Centro de Estudios de Física del Cosmos de Aragón, Plaza San Juan 1, 44001 Teruel, Spain}

\author{E. Lacruz}
\affiliation{Centro de Estudios de Física del Cosmos de Aragón, Plaza San Juan 1, 44001 Teruel, Spain}
\affiliation{Valencian International University,  Dept. of Mathematics, C. del Pintor Sorolla, 21, Ciutat Vella, 46002 Valencia, Spain}

\author[0000-0001-6097-5297]{R. Gonçalves}
\affiliation{Instituto Politécnico de Tomar, CI2 e U.D. Matemática e Física, 2300-313 Tomar, Portugal}

\author{B. Staels}
\affiliation{American Association of Variable Star Observers (AAVSO), 185 Alewife Brook Parkway, Suite 410, 02138 Cambridge, United States of America}

\author{R. Goossens}
\affiliation{Andromeda Vereniging voor Sterrenkunde en Ruimtevaart van de Dendervallei Denderweg 53, 9308 Aalst, Belgium}

\author{A. Henden}
\affiliation{American Association of Variable Star Observers (AAVSO), 185 Alewife Brook Parkway, Suite 410, 02138 Cambridge, United States of America}

\author{G. Walker}
\affiliation{Maria Mitchell Observatory, 3 Vestal Street, 02554 Nantucket, United States of America}

\author{J. A. Reyes}
\affiliation{Arroyo Observatory, Spain}

\author{S. Pastor}
\affiliation{Arroyo Observatory, Spain}

\author[0000-0002-9925-534X]{S. Kaspi}
\affiliation{School of Physics and Astronomy and Wise Observatory, Tel Aviv University, 6997801 Tel Aviv, Israel}

\author{M. Skrutskie}
\affiliation{Department of Astronomy, University of Virginia, Charlottesville, VA, United States of America}

\author[0000-0002-3323-9304]{A. J. Verbiscer}
\affiliation{Department of Astronomy, University of Virginia, Charlottesville, VA, United States of America}

\author{P. Martinez}
\affiliation{ADAGIO Association, Belesta Observatory (MPC A05), 550 route des étoiles, 31540 Toulouse, France}
\affiliation{International Occultation Timing Association/European Section, Am Brombeerhag 13, 30459 Hannover, Germany}

\author{P. André}
\affiliation{ADAGIO Association, Belesta Observatory (MPC A05), 550 route des étoiles, 31540 Toulouse, France}
\affiliation{International Occultation Timing Association/European Section, Am Brombeerhag 13, 30459 Hannover, Germany}

\author{J. L. Maestre}
\affiliation{Albox Observatory, Almería, Spain}

\author{F. J. Aceituno}
\affiliation{Centro Astronómico Hispano en Andalucía Observatorio de Calar Alto, Sierra de los Filabres, 04550 Gérgal, Almería, Spain}

\author[0000-0002-3105-7072]{P. Bacci}
\affiliation{Unione Astrofili Italiani (UAI), Italy}
\affiliation{Gruppo Astrofili Montagna Pistoiese 104, San Marcello Pistoiese, Italy}

\author{M. Maestripieri}
\affiliation{Unione Astrofili Italiani (UAI), Italy}
\affiliation{Gruppo Astrofili Montagna Pistoiese 104, San Marcello Pistoiese, Italy}

\author{M. D. Grazia}
\affiliation{Osservatorio Astronomico Montagna Pistoiese, San Marcello Pistoiese, Italy}

\author[0000-0003-2999-3563]{A. J. Castro-Tirado}
\affiliation{Instituto de Astrofísica de Andalucía (CSIC), Glorieta de la Astronomía s/n, 18008 Granada , Spain}
\affiliation{Unidad Asociada al CSIC, Departamento de Ingenier\'ia de Sistemas y
Autom\'atica, Escuela de Ingenier\'ias, Universidad de M\'alaga,
M\'alaga, Spain}

\author[0000-0002-7273-3671]{I. Pérez-Garcia}
\affiliation{Instituto de Astrofísica de Andalucía (CSIC), Glorieta de la Astronomía s/n, 18008 Granada , Spain}

\author[0009-0009-4604-9639]{E. J. Fernández García}
\affiliation{Instituto de Astrofísica de Andalucía (CSIC), Glorieta de la Astronomía s/n, 18008 Granada , Spain}

\author{E. Fernández}
\affiliation{Instituto de Astrofísica de Andalucía (CSIC), Glorieta de la Astronomía s/n, 18008 Granada , Spain}

\author{S. Messner}
\affiliation{H25 Harvest Moon Observatory, Northfield, MN, USA.}

\author{G. Scarfi}
\affiliation{K78 Astronomical Observatory Iota Scorpii of La Spezia, Italy}
\affiliation{International Occultation Timing Association/European Section, Am Brombeerhag 13, 30459 Hannover, Germany}

\author{H. Miku\v{z}}
\affiliation{Črni Vrh Observatory, Predgriže 29A, 5274 Črni Vrh nad Idrijo, Slovenia}
\affiliation{University of Ljubljana, Faculty of Mathematics and Physics, Jadranska 19, 1000 Ljubljana, Slovenia}

\author{J. Prat}
\affiliation{Observatorio Astronomico de Guirguillano, 31291 Guirguillano, Spain}

\author{P. Martorell}
\affiliation{Observatorio Astronomico de Guirguillano, 31291 Guirguillano, Spain}

\author[0000-0003-1149-3659]{D. Nardiello}
\affiliation{Dipartimento di Fisica e Astronomia ``Galileo Galilei" -- Università degli Studi di Padova, Vicolo dell'Osservatorio 3, I-35122 Padova, Italy}
\affiliation{National Astrophiscs Institute (INAF), Padova Astronomical Observatory, Vicolo dell'Osservatorio 3, I-35122 Padova, Italy}

\author[0000-0001-9770-1214]{V. Nascimbeni}
\affiliation{National Astrophiscs Institute (INAF), Padova Astronomical Observatory, Vicolo dell'Osservatorio 3, I-35122 Padova, Italy}

\author[0000-0002-4939-013X]{R. Sfair}
\affiliation{São Paulo State University (UNESP), School of Engineering and Sciences, 12516-410 Guaratinguetá, São Paulo, Brazil}
\affiliation{LESIA, Observatoire de Paris, Université PSL, CNRS, Sorbonne Université, 5 place Jules Jansse, 92190 Meudon,France}

\author{P. B. Siqueira}
\affiliation{São Paulo State University (UNESP), School of Engineering and Sciences, 12516-410 Guaratinguetá, São Paulo, Brazil}

\author{V. Lattari}
\affiliation{São Paulo State University (UNESP), School of Engineering and Sciences, 12516-410 Guaratinguetá, São Paulo, Brazil}

\author{L. Liberato}
\affiliation{Université Côte d’Azur, Observatoire de la Côte d’Azur, CNRS, Laboratoire Lagrange, Bd de l’Observatoire, CS 34229, 06304 Nice Cedex 4, France}

\author{T. F. L. L. Pinheiro}
\affiliation{São Paulo State University (UNESP), School of Engineering and Sciences, 12516-410 Guaratinguetá, São Paulo, Brazil}

\author[0000-0003-3194-5237]{T. de Santana}
\affiliation{São Paulo State University (UNESP), School of Engineering and Sciences, 12516-410 Guaratinguetá, São Paulo, Brazil}
\affiliation{Instituto Nacional de Pesquisas Espaciais (INPE), São José dos Campos, Brazil}

\author[0000-0003-1000-8113]{C. L. Pereira}
\affiliation{Observatório Nacional (ON/MCTI), R. General José Cristino 77, 20921 Rio de Janeiro, Brazil}
\affiliation{Laboratório Interinstitucional de e-Astronomia (LIneA/INCT do e-Universo), Av. Pastor Martin Luther King Jr, 126, 20765, Rio de Janeiro, Brazil}

\author{M.A. Alava-Amat}
\affiliation{Red Astronavarra Sarea, Astrosedetania, Spain}

\author{F. Ciabattari}
\affiliation{Monte Agliale Astronomical Observatory, Via Cune Motrone, 55023 Borgo a Mozzano, Italy}

\author{H. González-Rodriguez}
\affiliation{Observatorio de Forcarei, Galicia, Spain}

\author{C. Schnabel}
\affiliation{Agrupación Astronómica de Sabadell, Carrer Prat de la Riba, 116, 08206 Sabadell, Spain}
\affiliation{International Occultation Timing Association/European Section, Am Brombeerhag 13, 30459 Hannover, Germany}

\begin{abstract}

The physical and orbital parameters of Trans-Neptunian Objects (TNOs) provide valuable information about the Solar System's formation and evolution. In particular, the characterization of binaries provides insights into the formation mechanisms that may be playing a role at such large distances from the Sun. Studies show two distinct populations, and (38628) Huya occupies an intermediate position between the unequal-size binaries and those with components of roughly equal sizes. In this work, we predicted and observed three stellar occultation events by Huya. Huya and its satellite--S/2012~(38628)~1--were detected during occultations in March 2021 and again in June 2023.  Additionally, an attempt to detect Huya in February 2023 resulted in an additional single-chord detection of the secondary. A spherical body with a minimum diameter of D~=~165~km can explain the three single-chord observations and provide a lower limit for the satellite size. The astrometry of Huya's system, as derived from the occultations and supplemented by observations from the Hubble Space Telescope and Keck Observatory, provided constraints on the satellite orbit and the mass of the system. Therefore, assuming the secondary is in an equatorial orbit around the primary, the limb fitting was constrained by the satellite orbit position angle. The system density, calculated by summing the most precise measurement of Huya's volume to the spherical satellite average volume, is $\rho_{1}$~=~1073~$\pm$~66~kg~m$^{-3}$.  The density that the object would have assuming a Maclaurin equilibrium shape with a rotational period of 6.725~$\pm$~0.01~hours is $\rho_{2}$~=~768~$\pm$~42~kg~m$^{-3}$. This difference rules out the Maclaurin equilibrium assumption for the main body shape.

\end{abstract}



\section{Introduction}
\label{sec:Introduction}

The stellar occultation technique allows us to accurately measure an object using ground-based observations of a star from multiple stations. Our international collaboration has used this technique to derive physical properties of trans-Neptunian objects (TNOs) and Centaurs, enabling size and shape determination, detecting topographic features, and even discovering rings around these small Solar System objects \citep{Sicardy2011, Ortiz2012a,  Ortiz2017, Ortiz2020, Ortiz2023, Braga-Ribas2014, Braga-Ribas2023, Rommel2020, Rommel2023,  SantosSanz2021, Morgado2021, Morgado2023, Pereira2023}.

(38628) Huya was discovered from observations taken from Mérida-VEN \citep{Ferrin2000} and is a Neptune-crossing TNO located in the 2:3 mean motion resonance (MMR) with Neptune \citep{Gladman2008}, also known as a Plutino. The near-infrared spectra of Huya reveal i) evidence of methanol ice and ii) compatibility with the spectra of the binary Plutinos Mors-Somnus, 2007 JF$_{43}$, and Lempo \citep{Fornasier2013, Barkume2008, Mommert2012, SouzaFeliciano2018, SouzaFeliciano2024}. Observations from the Hubble Space Telescope (HST) revealed a $\approx$1.4-mag fainter companion located about 1740 km from the primary body, provisionally designated as S/2012 (38628) 1 \citep{Noll2012}, though its orbit remains unpublished.  Thermal measurements from the Herschel Space Observatory (HSO) and Spitzer Space Telescope (SST) allowed estimates of the area-equivalent diameters of Huya and its satellite of 406 $\pm$ 16 km and 213 $\pm$ 30 km, respectively \citep{Fornasier2013}. A multi-chord stellar occultation observed in March 2019 confirmed Huya's area-equivalent diameter of 411.0 $\pm$ 7.3 km \citep{SantosSanz2022}.

Binary systems among the Solar System's small bodies are thought to form through several possible mechanisms: capture, gravitational collapse, rotational fission, and giant impacts. Gravitational collapse tends to create binaries with nearly equal sizes and various separation distances \citep{Bernstein2023}. Rotational fission is particularly associated with the Haumea system \citep{Ortiz2012b}. In contrast, capture and giant impacts are more likely to result in larger TNOs with smaller satellites. Despite some exceptions \citep{Weaver2022}, a notable dichotomy has been observed in the relation between both component sizes and their separation distances: the largest known trans-Neptunian binaries (TNBs) generally have smaller satellites located within 100 times the radius of the primary body, while the smallest TNBs exhibit components of comparable size, with separations exceeding 100 times the radius of the primary body \citep{Bernstein2023}.  

\begin{figure}[!htb]
    \centering
    \includegraphics[width=0.8\linewidth]{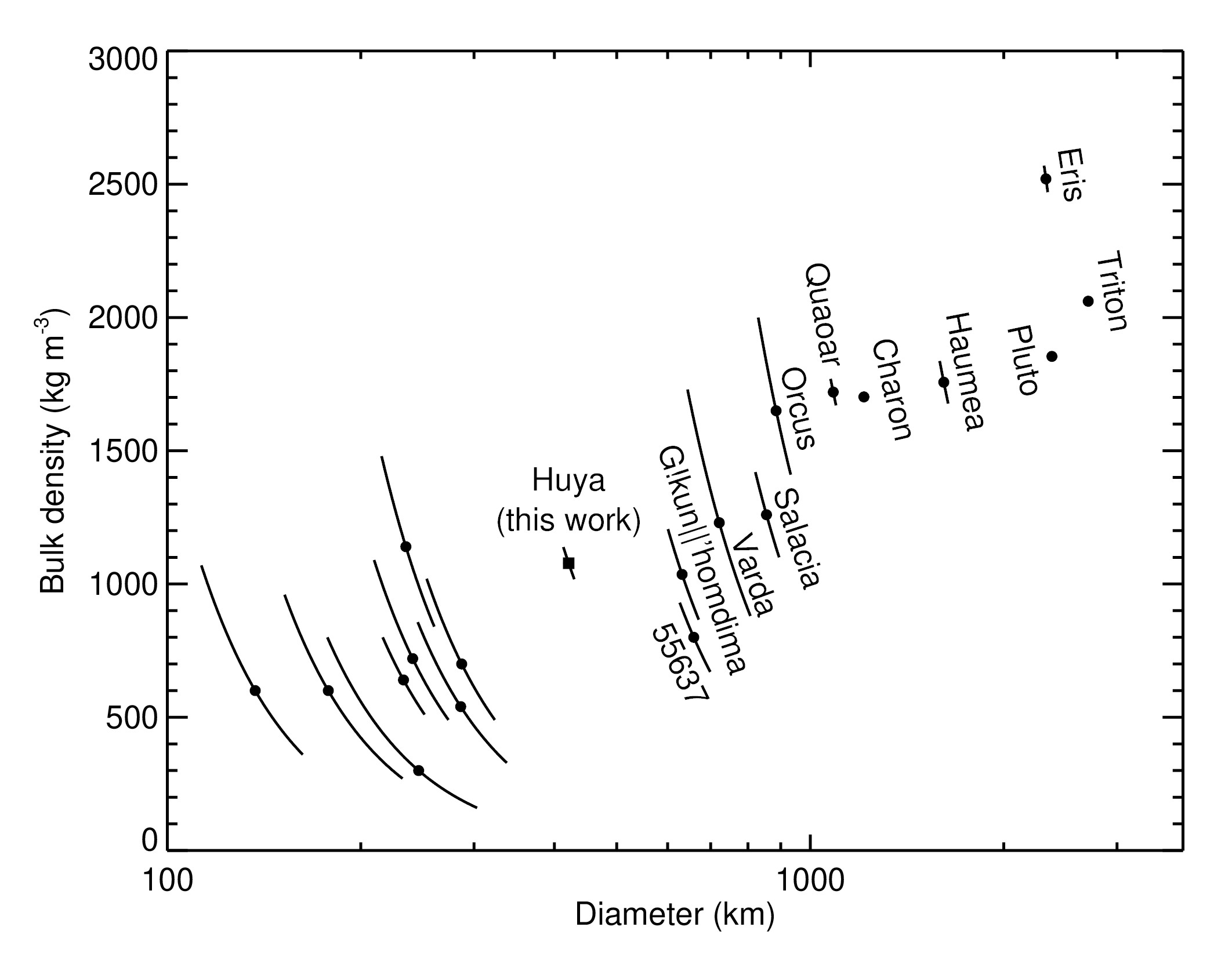}
    \caption{Objects' densities as a function of the diameter. The square marks the location of Huya using the published primary diameter \citep{SantosSanz2022} and the density derived in this work. Quaoar's density is from \cite{Kiss2024}. Other densities and primary diameters are from \cite{Grundy2019} and references therein.}
    \label{fig:grundy_adaptation}
\end{figure}

The Huya system occupies an intermediate position between the unequal-size binaries, characterized by a large primary and small moon, and those with components of roughly equal sizes \citep{Nesvorny2019}.  Despite the global density remaining unknown, an analysis based on the well-established primary diameter and the scheme proposed by \cite{Grundy2019} suggests that Huya is situated in an intermediate region between small, low-density binaries and large, high-density binaries (see Fig. \ref{fig:grundy_adaptation}). Thus, a comprehensive understanding of the physical and orbital properties of this object could provide valuable insights into the broader characteristics of the TNB population and the relationship between these two extreme binary populations.
\section{Prediction and observations}
\label{sec:Prediction}

Our international collaboration aims to make use of the accuracy offered by the stellar occultation technique to obtain the physical properties of TNOs and continuously improve our knowledge about the physical processes that take place in the outer Solar System region. Regular astrometric observations have been made since 2010 to update stars' positions and improve objects' ephemerides to accurately predict such events\footnote{More information is available at \url{https://lesia.obspm.fr/lucky-star/index.php}} by Centaurs and TNOs \citep{Ortiz2020}. Nowadays, thanks to the \textit{Gaia} stellar catalog releases \citep{Gaia2016,Gaia2018,Gaia2023}, only the small body ephemerides themselves require new astrometric observations to maintain accurate stellar occultation predictions. Huya's ephemeris includes astrometry obtained at ESO La Silla (Chile) in 2013, the Sierra Nevada Observatory (IAA/CSIC - Spain) from 2021 to 2023, and also the Pico dos Dias Observatory (OPD - Brazil) from 2017 to 2019\footnote{The OPD images were reduced with the astrometry tool from the Reduction of Astronomical Images Automatically \cite[\textsc{praia};][]{Assafin2023a}.}. 

\begin{figure}
    \gridline{\fig{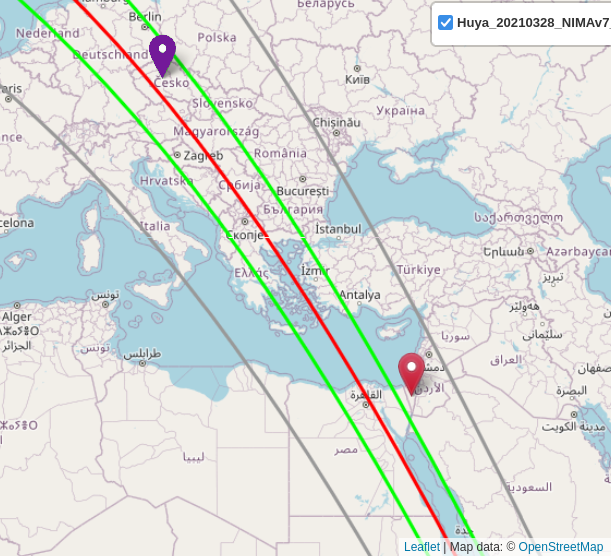}{0.45\textwidth}{(a){\label{fig:2021_03_prediction_map}}}
          \fig{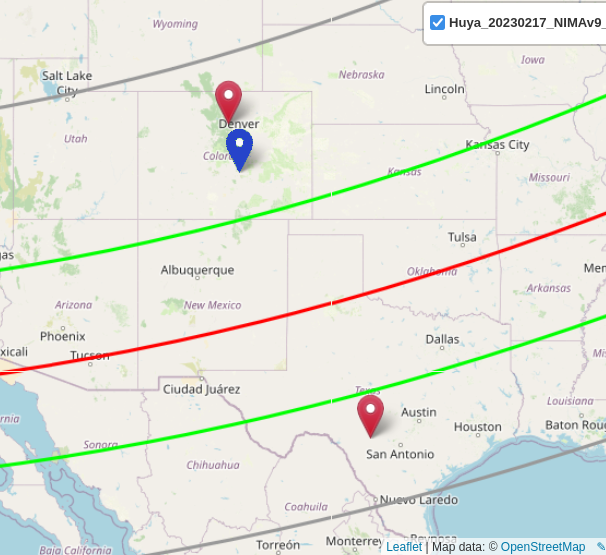}{0.45\textwidth}{(b)}{\label{fig:2023_02_prediction_map}}}
    \gridline{\fig{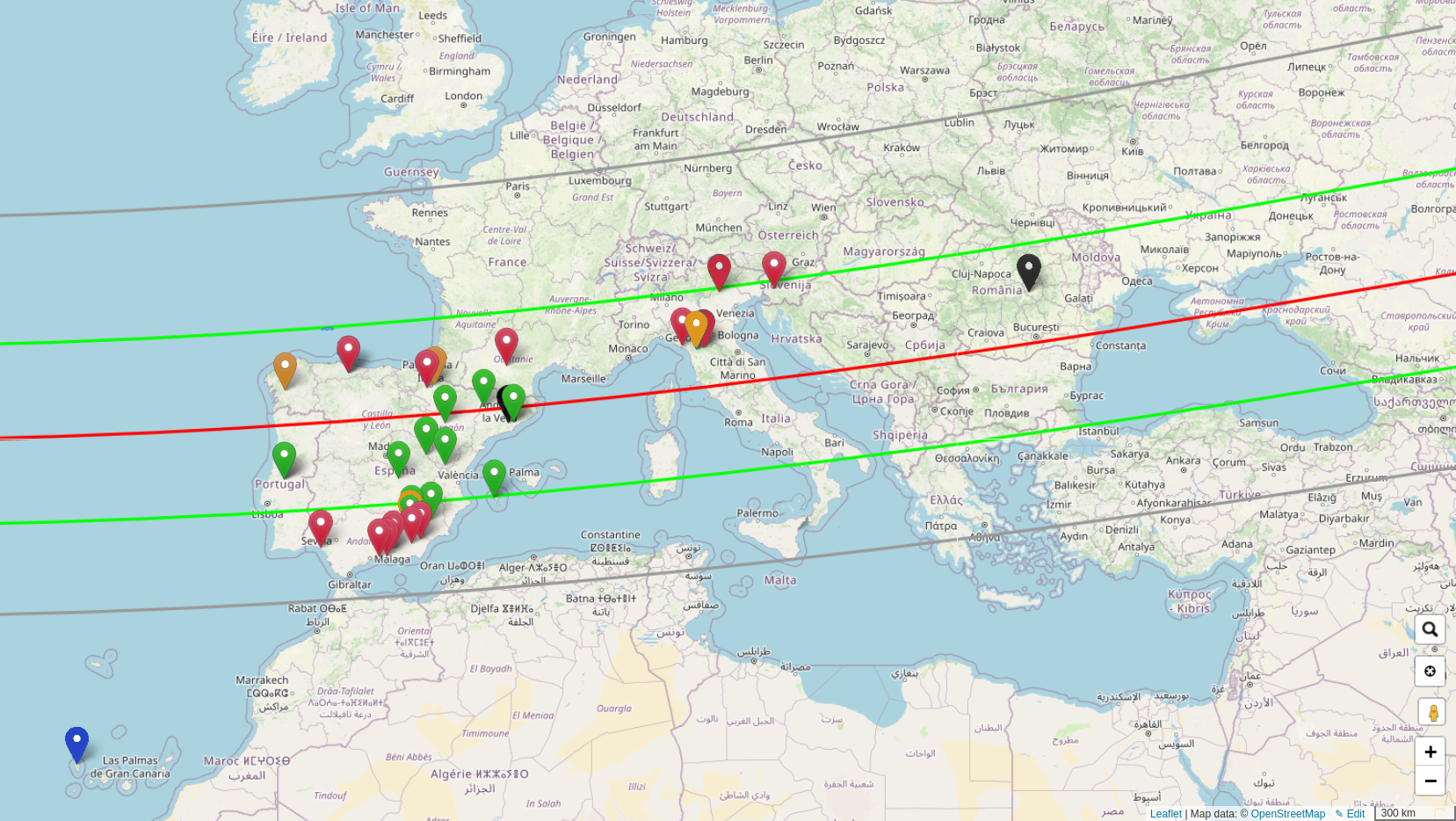}{0.95\textwidth}{(c)}{\label{fig:2023_06_prediction_map}}}
    \caption{Green lines present the predicted shadow path northern and southern limits with the $1\sigma$ uncertainties in gray for the stellar occultations recorded on (a) 2021-03-28, (b) 2023-02-17, and (c) 2023-06-24. The red line shows the predicted center for Huya's shadow path. The green markers depict the stations with positive detections of the primary component, the blue markers represent observatories that recorded only the satellite, and the observatory denoted by the purple marker in (a) recorded both components of the binary system. The red markers depict stations that recorded negative detections, inconclusive results are shown in black, and bad weather (or technical problems) are in orange. In panel c), the extensive distribution of observers over a large area reduces the ability to distinguish individual markers in regions with higher observers density. Maps were generated by the Occultation Portal web page described in \cite{Kilic2022}.}
    \label{fig:huya_general_view}
\end{figure}

Thanks to the improvement in Huya's orbit after the success of the 2019 occultation campaign \citep{SantosSanz2022}, we predicted and observed three additional stellar occultations by Huya. Predictions were performed using the Numerical Integration of the Motion of an Asteroid (\textsc{nima}; \citealt{Desmars2015}) and \textit{Gaia} stellar catalogs. The first event presented in this work occurred on 2021-03-28 when a stellar occultation (Fig. \ref{fig:huya_general_view}a) was detected from Ond\v{r}ejov observatory (Czech Republic), and a close negative chord was recorded from the Wise Observatory (Israel). Huya astrometry obtained from this single positive chord was used to improve the \textsc{nima} ephemeris and the prediction of future stellar occultation events. Later, a more exhaustive analysis of Ond\v{r}ejov data also revealed a stellar occultation by Huya's satellite (see Sect. \ref{sec:occ_satellite}). 

Almost two years later, on 2023-02-17, observers from the continental United States were contacted to attempt the stellar occultation of a V = 16.21 mag star by Huya.  Data were collected from three stations within the $1\sigma$ region of the predicted shadow path: one in the southern portion and two in the northern region of the shadow path (Fig. \ref{fig:huya_general_view}b). Among them, only the Penrose observatory recorded a positive. Due to technical issues, the image acquisition at Penrose could start only two seconds before the predicted time for the main body occultation for this station (11:42:01.00 UTC). According to \cite{Lindegren2021}, a well-behaved source from \textit{Gaia} stellar catalog will present a Renormalised Unit Weight Error (RUWE) $\approx$ 1. The target star has RUWE = 0.966 and no duplicate source flag. Additionally, the most recent \textsc{nima} v10 orbit fit for Huya has uncertainties of $\sigma_\alpha = 7$ milliarcseconds (mas) and $\sigma_\delta = 9$ mas at the time of the occultation. Even though, an offset of 62 seconds was observed between the predicted time and the center of the single-chord detection from Penrose, corresponding to $\approx$53 mas on the sky plane. Such an offset suggests that the detection should be attributed to the satellite rather than the main body (see discussion in Sect. \ref{sec:occ_satellite}).

The most recent stellar occultation by Huya occurred on 2023-06-24 and involved a V = 17.6 mag star. Observations were attempted from Portugal, Spain, Italy, and Slovenia, leading to the second multiple-chord event recorded for this object. A total of 30 stations participated in the observational campaign. Among these, 11 recorded positive detections by Huya, while one detected the satellite. Twelve data sets did not detect either components, two stations produced inconclusive results due to the low signal-to-noise ratio (S/N) of the target star in the images, and four stations were unable to acquire data due to bad weather conditions or technical failures (Fig. \ref{fig:huya_general_view}c). Table \ref{tab:target_stars} presents general information about the target stars, including their diameters at the object's geocentric distance (Star diam), calculated using the methods outlined in \cite{Rommel2023}, as well as the shadow velocity and the Fresnel effect in the occultation light curves, computed using the same approach as described in \cite{GomesJunior2022}.

\begin{table}[!htb]
    \caption{Occulted star designations and parameters at the closest approach instant (UTC) sorted by occultation date. The V magnitude was obtained from the NOMAD stellar catalog \citep{Zacharias2005}.}
    \label{tab:target_stars}
    \centering
    \begin{tabular}{c c c c c c c c}
    \hline \hline
    \textbf{Date} & \textbf{Gaia DR3 designation} & \textbf{Right ascension} & \textbf{Declination} & \textbf{V} & \textbf{Star diam} & \textbf{Fresnel} & \textbf{Velocity} \\
    \textbf{} &  & \textbf{(hh:mm:ss.ss)} & \textbf{($^\circ$ $'$ $''$)} & \textbf{(mag)} & \textbf{(km)} & \textbf{(km)} & \textbf{(km/s)} \\ \hline
    2021-03-28 & 4339984398716279808 & 17 02 24.10660 & -07 06 07.8921 & 17.60 & 0.12 & 1.21 & 8.85\\ 
    2023-02-17 & 4360090923037163136 & 17 22 02.20582 & -07 49 54.7261 & 16.21 & 0.36 & 1.23 & 19.27\\ 
    2023-06-24 & 4360429542557512064 & 17 16 43.65331 & -07 00 20.3449 & 17.60 & 0.18 & 1.20 & 22.43\\ \hline \hline
    \end{tabular}
\end{table}

Data sets were collected using a variety of telescope sizes, from compact 30-cm models to larger facilities such as the 1.5-meter telescope at Sierra Nevada Observatory (IAA/CSIC - Spain) and the La Palma 2-meter Liverpool telescope (IAC - Spain). Data quality varied with the exposure time and equipment used, but most observers did not use filters in order to increase the S/N. For time synchronization, among the received data, 10 used the Global Positioning System (GPS). The GPS antenna connects multiple atomic-clock-equipped GPS satellites and provides the Universal Time (UTC) with uncertainties of $\approx$ 5 nanoseconds \citep{Gamage2024}. Most remaining stations relied on the Network Time Protocol (NTP) to synchronize image timestamps to universal time. However, NTP reliance on internet connectivity introduces some known issues such as network congestion and clock drift \citep{Gamage2024}. Consequently, NTP sync data must be handled with caution to account for these uncertainties. Following the observations, data sets, and reports were uploaded to the Occultation Portal platform\footnote{\url{https://occultation.tug.tubitak.gov.tr/}} \citep{Kilic2022}. Details of all participating observers and their instruments are provided in Appendix \ref{sec:appendixd}.

\section{Data analysis and results}
\label{sec:data_analysis}
This section provides a comprehensive overview of the data reduction and analysis of stellar occultations, along with the determination of the satellite's orbit. It also includes a detailed presentation of the results obtained in this work.

\subsection{Occultations by Huya}
The data sets from the stellar occultations described in this paper have a variety of formats and image quality. When FITS-format and calibration images were obtained, they underwent bias, dark (when necessary), and flat field corrections using standard procedures implemented in the \textsc{ccdproc v2.4.1} Python library\footnote{Documentation available on \url{https://ccdproc.readthedocs.io/en/latest/}} \citep{Craig2023}. The \textit{avi} and \textit{ser} video files were first converted to FITS format using a script based on the \textsc{opencv-python} v4.7.0.72 library\footnote{Documentation available on \url{https://pypi.org/project/opencv-python/}} \citep{opencv_library}. This script extracts the odd and even fields from the video file and combines them to obtain one full video frame, a required procedure when dealing with interlaced video data\footnote{See a detailed explanation about camera's video modes here: \url{http://www.dangl.at/ausruest/vid_tim/vid_tim1.htm\#wat910hxeia}}. Depending on the CCD camera acquisition mode, the extracted frames are repeated copies of the same exposure. Therefore, the frames were processed using a Python script based on \textsc{astropy} v5.2.1 \citep{Astropy2022} and stacked using each pixel's median flux value to mitigate the effect of electronic noise in the individual copies. The number of stacked frames depends on the camera model and the used acquisition mode. Due to the way that the acquisition software writes the time over the frames, offsets are also required in some instrument configurations. For instance, in this work, the Sabadell data set was obtained with a WATEC 910HX camera set to the CCIR-x256 mode. In this acquisition mode, the extracted frames must be stacked every 127, and a time correction of -2.54 seconds is required to recover the correct UTC information\footnote{See the note in table WAT-910  (CCIR) here: \url{http://www.dangl.at/ausruest/vid_tim/vid_tim1.htm\#wat910hxeia}}. We carefully check that no frames were lost, as this could lead to the mixing of frames from different exposures. In addition, as time is written over the frames, the resulting images do not contain time information in the header. Therefore, the exposure time and the timestamp of the first image must be manually provided to the photometry software to properly generate the occultation light curve.

Relative aperture photometry was done using the photometry tool from the Package for the Reduction of Astronomical Images Automatically \cite[\textsc{praia};][]{Assafin2023b}, with aperture sizes optimized to maximize the star's S/N. The background-corrected flux of the target star was divided by the unweighted average fluxes of the calibration stars to remove the signature of atmospheric variability. A polynomial function was used to flatten the light curve, which was then divided by its average to normalize the flux ratio to unity outside the occultation\footnote{A detailed description of the procedures involving the light curve process by \textsc{praia} photometry task can be obtained from the user guide here: \url{https://ov.ufrj.br/praia-photometry-task/}}. Based on the star and object magnitudes, a maximum brightness contribution of $14.8\%$ was expected from the Huya system during the stellar occultations observed in March 2021 and June 2023. Consequently, the fluxes during the occultation were normalized to this value. The expected brightness contribution from the Huya system for the February 2023 event is only $4\%$ and, given the data dispersion, can be neglected. 

\begin{figure}
    \gridline{\fig{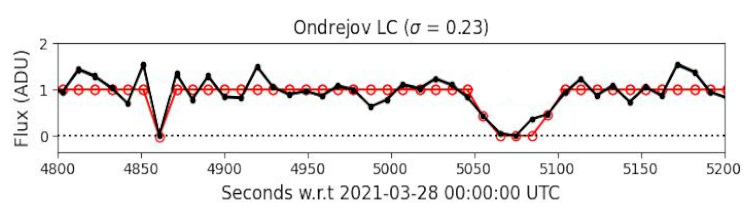}{0.499\textwidth}{(a)}{\label{fig:ondrejov_lc}}
          \fig{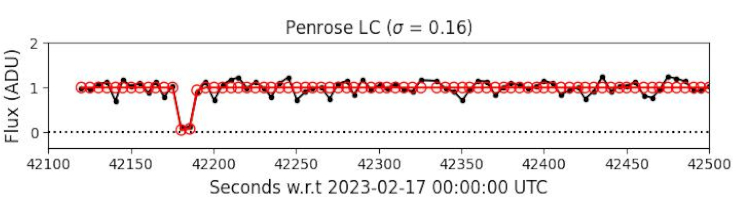}{0.5\textwidth}{(b)}{\label{fig:penrose_lc}}}
    \gridline{\fig{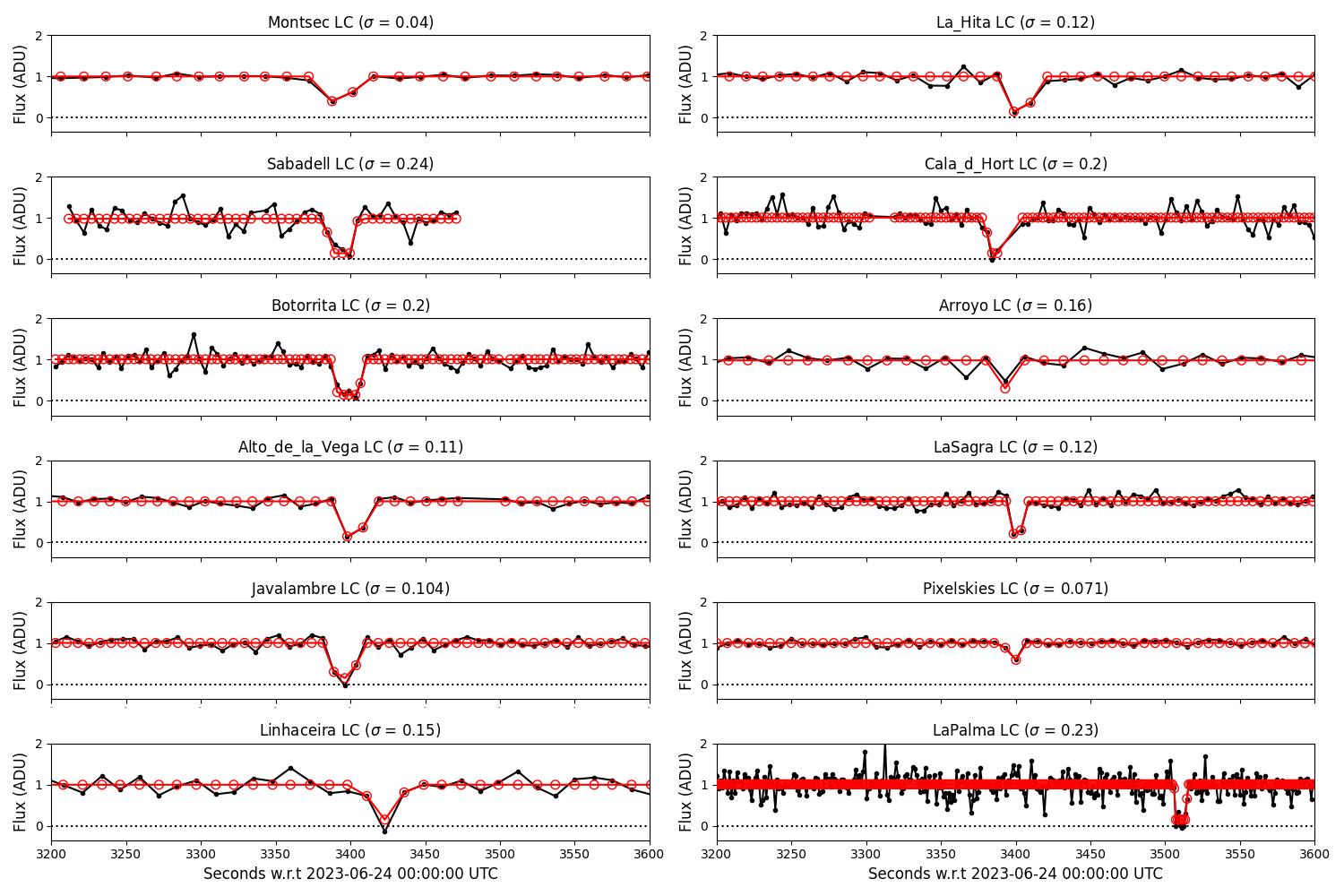}{\textwidth}{(c)}{\label{fig:lapalma_lc}}}
    \caption{Occultation light curves of Huya system as recorded on (a) 2021-03-28, (b) 2023-02-17, and (c) 2023-06-24. Black dots and lines represent the observed data. The red dotted lines show the synthetic light curve model (see text).\label{fig:huya_lcs}}
\end{figure}

The ingress and egress instants were derived using the Stellar Occultation Reduction and Analysis package, v0.3.1\footnote{Documentation available on \url{https://sora.readthedocs.io/latest/}} \cite[\textsc{sora};][]{GomesJunior2022}. These instants were determined by modeling the positive light curves with a sharp-edge model, which was convolved with the stellar diameter at the object's distance, Fresnel diffraction effects, finite exposure time, and the CCD bandwidth. Since most data sets were acquired without filters unless otherwise specified in Table \ref{tab:observational_circumstances_positive}, the wavelength range used for Fresnel diffraction calculations was $\lambda$ = 700 $\pm$ 300 nm. The sub-kilometer effects of the stellar diameter and Fresnel diffraction on the light curve models (Table \ref{tab:target_stars}) are negligible considering the shortest exposure times for each recorded event (71 km, 96 km, and 18 km for 2021-03-28, 2023-02-17, and 2023-06-25, respectively). All positive light curves and their synthetic models derived using \textsc{sora} v0.3.1 are presented in Figure \ref{fig:huya_lcs}. The ingress and egress instants, along with the $1\sigma$ uncertainties, are summarized in Table \ref{tab:times}.

\begin{table}[!htb]
\centering
\caption{Ingress and egress instants (UTC) with $1\sigma$ uncertainties for each positive detection of the stellar occultation events presented in this work.}
    \label{tab:times}
    \begin{tabular}{l c c c c}\hline \hline
                       &\multicolumn{2}{c}{\textbf{Huya times}}&\multicolumn{2}{c}{\textbf{Satellite times}}\\
    \textbf{Sites}     &  \textbf{Ingress}         & \textbf{Egress}           & \textbf{Ingress}         & \textbf{Egress}\\
                       & (hh:mm:ss.ss $\pm$ ss.ss) & (hh:mm:ss.ss $\pm$ ss.ss) & (hh:mm:ss.ss $\pm$ ss.ss) & (hh:mm:ss.ss $\pm$ ss.ss)\\ \hline \hline
    \multicolumn{5}{c}{\textbf{2021 March 28}} \\  
    Ond\v{r}ejov           & 01:24:14.7 $\pm$ 1.8      & 01:24:54.5 $\pm$ 1.7      & 01:20:56.9 $\pm$ 2.0 & 01:21:05.4 $\pm$ 2.5\\ \hline
    \multicolumn{5}{c}{\textbf{2023 February 17}} \\ 
    Penrose            & -                         & -                         & 11:42:58.43 $\pm$ 0.90 & 11:43:07.60 $\pm$ 0.90\\ \hline
    \multicolumn{5}{c}{\textbf{2023 June 24}} \\ 
    Montsec            & 00:56:25.30 $\pm$ 0.54    & 00:56:40.90 $\pm$ 0.55    & - & -\\  
    Botorrita          & 00:56:29.5 $\pm$ 1.7      & 00:56:47.8 $\pm$ 1.1      & - & -\\ 
    Sabadell           & 00:56:21.2 $\pm$ 2.1      & 00:56:37.5 $\pm$ 1.3      & - & -\\ 
    Alto de la Vega    & 00:56:32.8 $\pm$ 1.2      & 00:56:51.2 $\pm$ 1.4      & - & -\\ 
    Javalambre         & 00:56:26.4 $\pm$ 1.0      & 00:56:44.7 $\pm$ 0.85     & - & -\\ 
    Linhaceira         & 00:56:55.8 $\pm$ 2.2      & 00:57:11.9 $\pm$ 2.4      & - & -\\ 
    La Hita            & 00:56:33.4 $\pm$ 2.4      & 00:56:54.6 $\pm$ 2.4      & - & -\\ 
    Cala d' Hort       & 00:56:20.3 $\pm$ 1.2      & 00:56:35.4 $\pm$ 7.8      & - & -\\ 
    Arroyo             & 00:56:28.9 $\pm$ 5.0      & 00:56:37.0 $\pm$ 5.0      & - & -\\ 
    La Sagra           & 00:56:36.5 $\pm$ 0.6      & 00:56:45.6 $\pm$ 0.9      & - & -\\ 
    Pixelskies         & 00:56:39.9 $\pm$ 1.7      & 00:56:44.7 $\pm$ 1.5      & - & -\\
    La Palma           &       -                   &   -                       & 00:58:26.42 $\pm$ 0.29 & 00:58:34.43 $\pm$ 0.3\\ \hline \hline
    \end{tabular}
\end{table}
The derived ingress and egress instants, along with their $1\sigma$ uncertainties, are then projected onto the sky plane (see equations 7-9 from \cite{GomesJunior2022}). The limb-fitting procedure depends on the number of data points available and involves minimizing the classical $\chi^2$ per degree of freedom ($\chi^2_{pdf}$) function. A satisfactory solution is indicated by $\chi^2_{pdf}=\chi^2/(N-M) \approx 1$, where $N$ is the number of data points and $M$ is the number of fitted parameters. Among the three events presented here, only the multi-chord detection of Huya in June 2023 provides sufficient data ($N$ $>$ 5) for an ellipse model ($M$ = 5) to be fitted. The limb-fitting started with a general fit using only the Montsec, Sabadell, Botorrita, and La Sagra GPS data sets. This preliminary fit, filtered by the Belesta close negative, was used to derive the normal to the object's surface velocities in each positive chord extremities \citep{GomesJunior2022}. The average of both values is then used to recalculate the instants, leading to the values presented in Table \ref{tab:times}. 

\begin{figure}
    \gridline{\fig{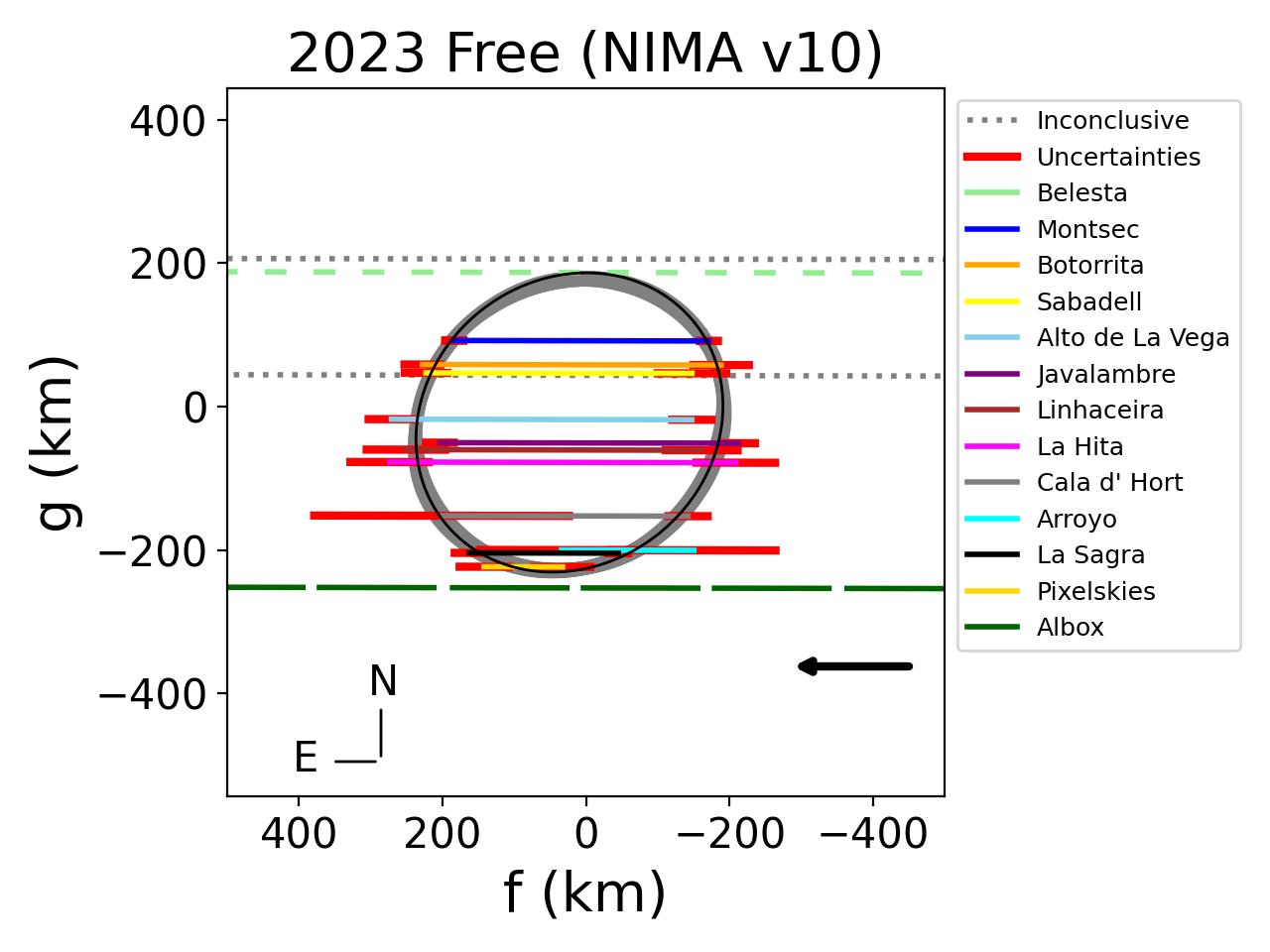}{0.5\textwidth}{(a)}{\label{fig:june2023_ellipse_fit_main_free}}
          \fig{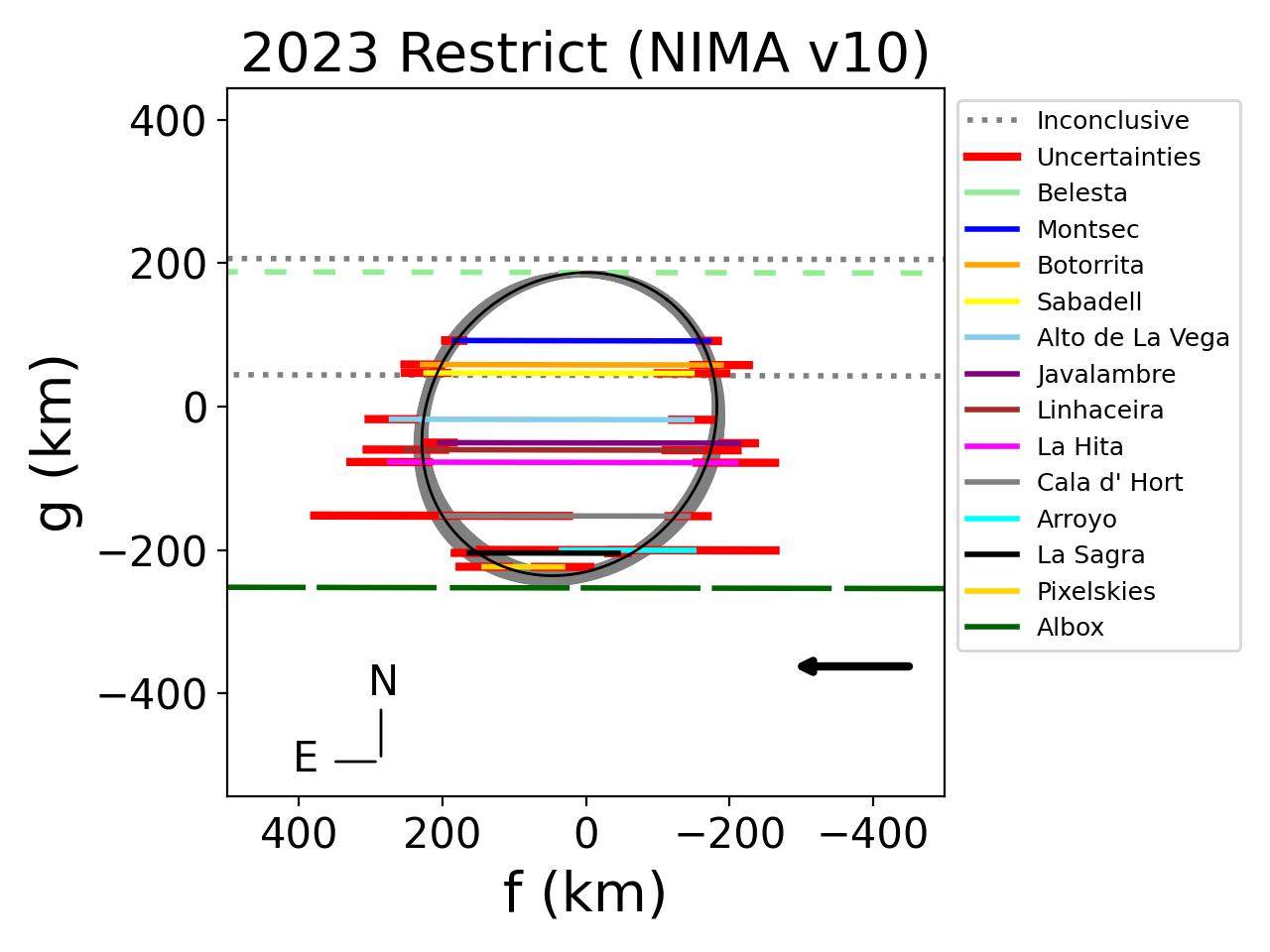}{0.5\textwidth}{(b)}{\label{fig:june2023_ellipse_fit_main_restrict}}}
    \caption{Stellar occultation by Huya on 2023-06-24 with the location of the stations that had inconclusive results indicated by gray dotted lines, the positive detections represented by colorful solid lines, and the red segments denoting the $1\sigma$ uncertainties. The light green segments represent the exposure times recorded at the Belesta Observatory, corresponding to the closest negative chord to the north of the observed profile. The dark green dashes indicate the exposure times captured at the Albox station, which corresponds to the closest negative chord to the south of the observed limb. The white space between the dashes reflects the readout time between exposures at both stations. The black ellipse is the best limb solution using the a) \textit{2023 Free} and b) \textit{2023 Restrict} approaches, respectively. The gray region represents the $1\sigma$ uncertainty for each limb determination approach. The direction of the shadow's movement is indicated by the black arrow.}
    \label{fig:Huya_limb_fittings}
\end{figure}

Huya's limb from the June 2023 stellar occultation was determined using all positive chords, excluding the La Palma data set associated with the satellite. The ellipse fit provides the center of the observed profile ($f$, $g$) regarding the object's ephemeris, the apparent semi-major and semi-minor axes ($a'$, $b'$), the position angle (PA) of the semi-minor axis regarding the celestial North and, the object's apparent oblateness ($\epsilon'$). This analysis, involving $N$ = 22 data points, used two distinct methods: (i) a free search of the five ellipse parameters ($f$, $g$, $a'$, $\epsilon'$ and PA) and (ii) a constrained search based on position angles within PA = $53.7^\circ\pm 2.2^\circ$. The position angle interval for the constrained search was determined from the satellite orbit data (to be discussed in Sect. \ref{sec:satelite_orbit}), under the assumption that Huya has an oblate shape and that the secondary orbits the primary along its equatorial plane. In both methods, limb solutions intersecting with negative data recorded from Belesta were excluded using the SORA v0.3.1 $filter\_negative\_chord$ function (Fig. \ref{fig:Huya_limb_fittings}). The results of both limb searches are presented in the first and second columns of Table \ref{tab:fitted parameters}, where $R_{equiv}$ is the obtained area-equivalent radius (km) and the $R_{dispersion}$ corresponds to the radial residuals (km) between best-fitted ellipse and observed data points. The \textit{2023 Restrict} solution was used to obtain Huya's limb from the single detection acquired in 2021. The last column presents the results obtained using the same assumptions for the constrained limb fitting procedure as in the \textit{2023 Restrict} approach, but using a position angle range of PA = 51.6$^\circ$ $\pm$ 2.2$^\circ$ applied to the 19 positive chords published in 2019 (Fig. \ref{fig:2019_limb_restrict}). The large $\chi^2_{pdf}$ obtained for \textit{2019 Restrict} solution (Appendix \ref{sec:appendix_2019}) may suggest that some of the previous published positive chords need time offsets or that Huya limb presented some topography at that event. The putative topographic features observed in the 2019 data but not present in the June 2023 records can be explained by a combination of larger error bars in the 2023 data and the changes in the rotational phase during which the 2023 observation was made. Also, we cannot discard the possibility of a big feature in the northern part of the object profile, causing the Belesta data set to be a negative chord.

\begin{table}[!htb]
    \centering
    \caption{Parameters of Huya's best-fitted limb solutions ($1\sigma$) derived for each approach (see text). The $^\ast$ symbol marks the position angle values used to constrain the limb solutions.}
    \label{tab:fitted parameters}
    \begin{tabular}{c c c c}\hline \hline
            \textbf{Parameter}&\textbf{2023 Free} & \textbf{2023 Restrict} & \textbf{2019 Restrict}\\ \hline
            $f$               &  22.3 $\pm$ 6.7 km     & 22.2 $\pm$ 6.6 km        & 49.9 $\pm$ 0.15 km\\
            $g$               &  -24.9 $\pm$ 7.9 km    & -25.7 $\pm$ 5.0 km       & 26.61 $\pm$ 0.08 km\\
            $a'$              & 222.5 $\pm$ 9.1 km     & 218.7 $\pm$ 8.1 km       & 218.05 $\pm$ 0.11 km\\
            $b'$              & 198.7 $\pm$ 15.2 km    & 200.3 $\pm$ 14.9 km      & 195.59 $\pm$ 0.24 km \\
            $\epsilon'$       & 0.107 $\pm$ 0.058      & 0.084 $\pm$ 0.059        & 0.103 $\pm$ 0.001\\
            PA                & 31.5$^\circ$ $\pm$ 15.5$^\circ$& $^\ast$53.7$^\circ$ $\pm$ 2.2$^\circ$& $^\ast$51.6$^\circ$ $\pm$ 2.2$^\circ$\\ 
            R$_{\rm equiv}$       & 210.3 $\pm$ 11.0 km & 209.3 $\pm$ 10.3 km     & 206.5 $\pm$ 0.16 km\\
            $R_{\rm dispersion}$  & 3.6 $\pm$ 25.5 km   & 5.0 $\pm$ 26.0 km       & 1.1 $\pm$ 11.6 km\\ 
            $\chi^2_{\rm pdf}$    & 0.682               & 0.745                   & 210.9\\ \hline \hline
    \end{tabular}
\end{table}

Huya's global density can be determined through two different methods: i) using the system volume from the occultations and the mass derived from the satellite orbit fit presented in this work (Table \ref{tab:binary_orbit}) or; ii) assuming a Maclaurin equilibrium shape and taking into account Huya's rotational period. The first approach uses the fundamental equation for the density ($\rho_1 = M/V$), where $M$ is Huya's system mass and $V$ denotes the system's total volume. Huya's volume was obtained from the assumption of an oblate spheroid shape with true axes $a=b=a'=218.05\pm0.11$ km and $c=a(1-\epsilon)=187.5\pm2.4$ km, where $\epsilon$ is the true oblateness considering the equivalent radius and aspect angle (see Eq. \ref{eq:epsilon_true} in Appendix \ref{sec:appendix_density}). The determination of the satellite volume is under the assumption of a spherical body with a diameter ranging from the minimum obtained from the stellar occultation single-chords D = 165 km to the maximum value from the thermal D = 243 km (see Appendix \ref{sec:appendix_density}). As a result, we obtained a system density of $\rho_1$ = 1073 $\pm$ 66 kg~m$^{-3}$.

The second method assumes a Maclaurin equilibrium shape for the primary and uses the following equation, as in \cite{Sicardy2011,Braga-Ribas2013}:
\begin{equation}
    \rho_2 = \frac{4\pi}{P^2G}\ \frac{sin^2(\theta)tan(\theta)}{2\theta[2+cos(2\theta)]-3sin(2\theta)}
    \label{eq:density}
\end{equation}
where $\theta$ is Huya's aspect angle, which was assumed to be the same as the satellite orbit opening angle of $\theta = 60.0^\circ \pm 3.5^\circ$ for the June 2023 stellar occultation. $G$ is the gravitational constant, and $P$ is the published rotational period of 6.725 hours \citep{SantosSanz2022}. As the rotational period uncertainties were not given, we assumed an error of 0.01 hours. As a result, we obtained a density  $\rho_2$ = 768 $\pm$ 42 kg~m$^{-3}$, where uncertainty comes from the classic uncertainty propagation formula. The discrepancy between both density values is discussed in Sect. \ref{sec:discussion}.

\subsection{Occultations by Huya's satellite}
\label{sec:occ_satellite}

\begin{figure}[!htb]
    \gridline{\fig{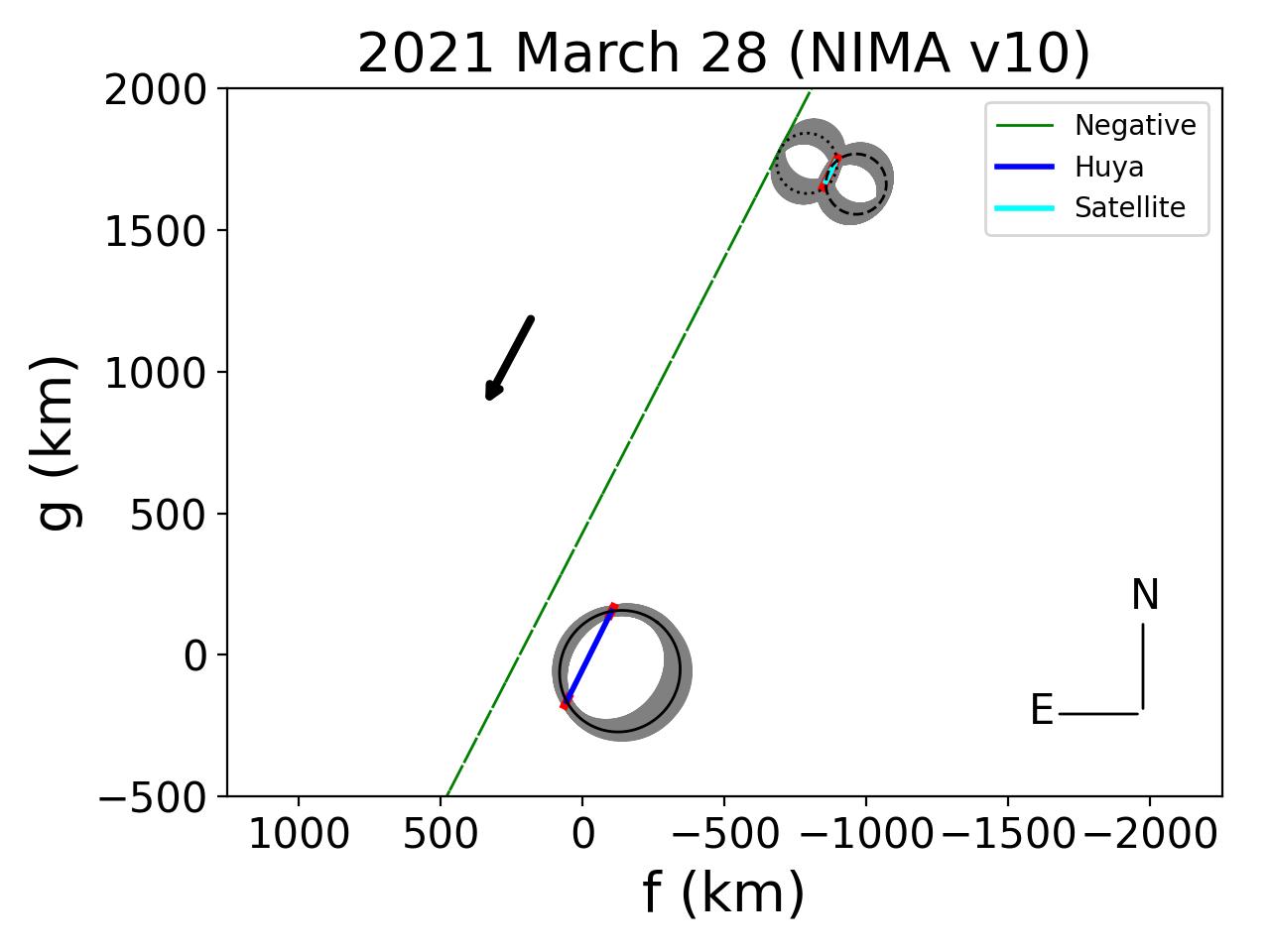}{0.5\textwidth}{(a)}{\label{fig:huya_2021}}
          \fig{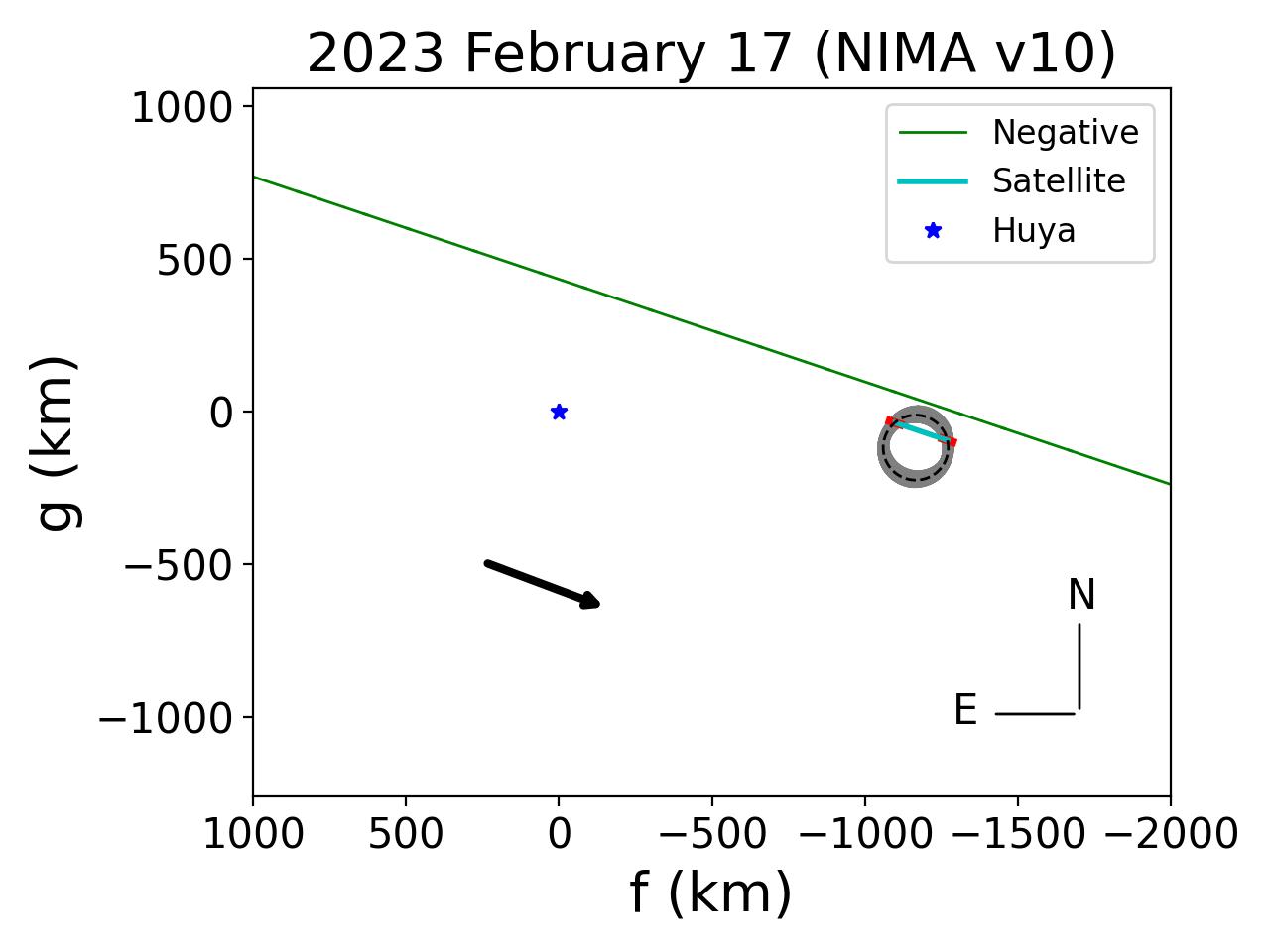}{0.5\textwidth}{(b)}{\label{fig:huya_feb2023}}}
    \gridline{\fig{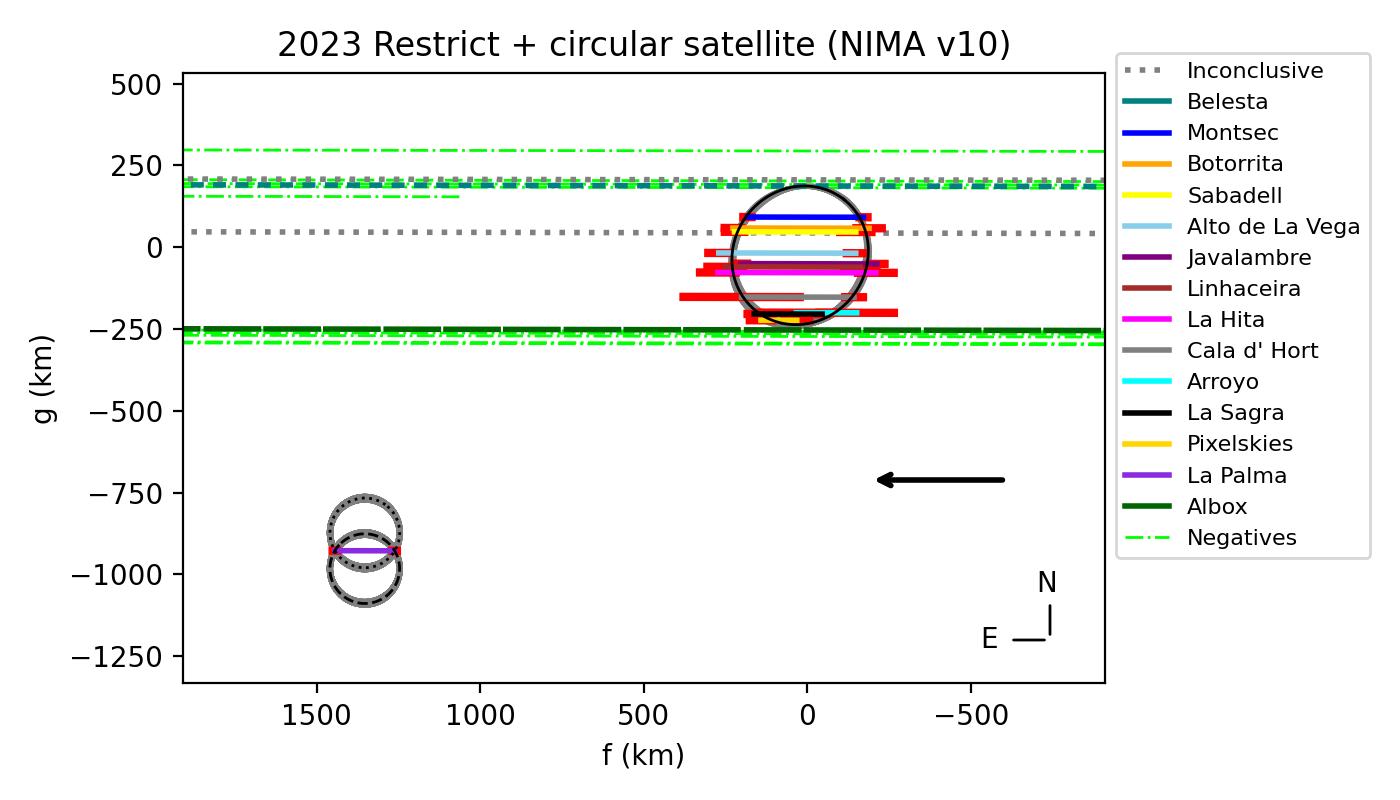}{\textwidth}{(c)}{\label{fig:june2023_ellipse_fit}}}
    \caption{Stellar occultations by the Huya system observed on (a) 2021-03-28, (b) 2023-02-17, and (c) 2023-06-24, as analyzed in this study. Positive chords are presented by solid lines in colors with error bars in red segments. Close negative exposures are presented by darker green segments, and other negative data sets are presented in light-green dash-dot lines. Inconclusive results are represented by dotted gray lines. The blue star marks the Huya predicted position on 2023 February 17. The black ellipse is the best solution for the primary, and the shadow region represents the $1\sigma$ uncertainty. Dotted and dashed circles show the solutions for the satellite detections (see text). The black arrow shows the direction of the shadow movement.}
    \label{fig:limb_fittings}
\end{figure}

This study presents three single-chord detections of S/2012 (38628) 1, hereafter referred to as Huya's satellite. Single-chord detections do not allow for a complete limb fitting. Consequently, we assumed a circular limb with the published radius of 106.5 km \citep{Fornasier2013} and allowed the center ($f$, $g$) to vary to obtain the satellite astrometry.  The Ond\v{r}ejov light curve, acquired in March 2021, has a notable standard deviation (0.23) and an exposure time of 8.0$\ $s. Despite this, a 4.3$\sigma$ drop in the stellar flux was identified before the occultation by the main body (Fig. \ref{fig:huya_lcs}a). The satellite positive chord has a length of 73 $\pm$ 40 km and is positioned at 1910 $\pm$ 55 km northwest of Huya's projected center. The negative data set acquired from Israel for the same event does not provide substantial constraints for the satellite circular limb solutions, leading to two astrometric solutions (Figure \ref{fig:limb_fittings}a, Table \ref{tab:astrometry_results2}). The most recent satellite detection happened on 2023 June 24, from the La Palma observatory (Fig. \ref{fig:huya_lcs}c), with an average separation of 1603 km from the primary in the southeast direction. This record represents the most precise measurement of Huya's satellite limb, yielding a chord length of 179 $\pm$ 14 km. However, despite the availability of many negative data sets for this event, none of the negatives are close enough to the La Palma observatory in order to constrain the satellite limb solution. As a result, two equally plausible solutions are obtained, as shown in Figure \ref{fig:limb_fittings}c and Table \ref{tab:astrometry_results2}.

\pagebreak

The single-chord data acquired from Penrose observatory in February 2023 revealed a nine-second drop in the stellar flux (Fig. \ref{fig:huya_lcs}b), which corresponds to a chord of $177\ \pm\ 35$ km on the sky plane. This was the only positive data for this event, with the observer positioned to the north of the predicted shadow path within the $1\sigma$ region. The data acquisition from Penrose observatory began only 2 seconds before, and the positive chord was recorded 62 seconds after the predicted instant for the Huya occultation at that location (11:42:01.00 UTC). \textit{Gaia} catalog provides a no duplicate flag and a RUWE = 0.996 for the target star, where RUWE = 1 corresponds to a perfectly well-behaved source. Additionally, the \textsc{nima} v10 orbit's precision at the event's date has uncertainties below 10 mas in both coordinates. Therefore, the offset of this positive detection cannot be attributed to bad stellar astrometry or large uncertainties in the ephemeris. It is more than $5\sigma$ away from Huya's predicted position, and such a large astrometric offset strongly suggests that the occultation was caused by Huya's satellite rather than the primary body. Therefore, a circular limb was fitted to this positive detection, and the solutions were filtered by the close negative recorded at the Nederland observatory, providing a single-center solution (Figure \ref{fig:limb_fittings}b, Table \ref{tab:astrometry_results2}). Since Huya was not detected during this event, its predicted position by \textsc{nima} v10 (with uncertainties) was used to calculate the relative position presented in Table \ref{tab:binary_astrometry} and the separation of 1173 $\pm$ 150 km between both components on the sky plane.  

\begin{table}[!htb]
    \centering
    \caption{Astrometry for Huya's system derived from the three stellar occultation events. The relative astrometry between Huya's satellite and the main body is presented in Table \ref{tab:binary_astrometry}.}
    \label{tab:astrometry_results2}
    \begin{tabular}{l l l l l} \hline \hline
         \textbf{Object}      & \textbf{Date}                          & \textbf{Right ascension ($\alpha$)}    & \textbf{Declination ($\delta$)}       & Solution  \\
                              & (yyyy-mm-dd hh:mm:ss.ss)               & (hh mm ss.ss $\pm$ mas)     & ($^\circ$ $'$ $''$ $\pm$ mas) &  \\ \hline
                              & 2019-03-18 00:43:28.44                 & 16 41 06.419830 $\pm$ 0.11  & -06 43 34.58532 $\pm$ 0.14 & 2019 Restrict\\ \cline{2-5}
         \multirow{2}{*}{Huya}& 2021-03-28 01:13:48.10                 & 17 02 24.10662 $\pm$ 1.6    & -07 06 07.8917 $\pm$ 1.3 & -\\ \cline{2-5}
                              & \multirow{2}{*}{2023-06-24 00:58:10.9 }& 17 16 43.653343 $\pm$ 0.52  & -07 00 20.11832 $\pm$ 0.78 & 2023 Free\\ \cline{3-5}
                              &                                        & 17 16 43.653342 $\pm$ 0.51  & -07 00 20.11836 $\pm$ 0.72 & 2023 Restrict\\\hline
        \multirow{5}{*}{Satellite}& \multirow{2}{*}{2021-03-28 01:13:48.10}& 17 02 24.10445 $\pm$ 1.1& -07 06 07.8046 $\pm$ 2.1  & Southern\\ 
                              &                                        & 17 02 24.10391 $\pm$ 1.3    & -07 06 07.8084 $\pm$ 1.8  & Northern\\ \cline{2-5}
                              & 2023-02-17 11:43:38.26                 & 17 22 02.197179 $\pm$ 0.75  & -07 49 54.5104 $\pm$ 1.2    & Northern \\ \cline{2-5}
                              & \multirow{2}{*}{2023-06-24  00:58:10.9}& 17 16 43.657737 $\pm$ 0.48  & -07 00 20.16006 $\pm$ 0.82 & Southern \\ 
                              &                                        & 17 16 43.657738 $\pm$ 0.48  & -07 00 20.16537 $\pm$ 0.82 & Northern \\ \hline \hline
    \end{tabular}
\end{table}


\subsection{Satellite orbit determination}
\label{sec:satelite_orbit}
The detection of Huya's satellite during three separate occultations, along with resolved images of the system from the Hubble Space Telescope (HST) and Keck Observatory, has enabled the determination of the satellite's orbit. HST archival images were taken by Programs 9110 and 12468, using the Space Telescope Imaging Spectrograph (STIS) in 2002 and the Wide Field Camera 3 (WFC3) in 2012. STIS images were acquired with no filter, while WFC3 images were taken with the F606W and F814W filters. Relative astrometry from these images was extracted using Point Spread Function (PSF) fitting techniques, employing model PSFs from TinyTim in a well-validated processing pipeline \citep[e.g.,][]{grundy200842355, grundy2009mutual}. All HST images used in this work are available on Mikulski Archive for Space Telescopes (MAST): \dataset[10.17909/8krd-3h13]{http://dx.doi.org/10.17909/8krd-3h13} 
Additionally, two observations were obtained in 2021 using the narrow camera of NIRC2 and the laser guide star adaptive optics \citep[LGS AO;][]{wizinowich2006} at the Keck Observatory. These observations, made with the infrared $H$ filter (wavelengths from $\sim$1.48 to 1.77 $\mu$m), involved dithered exposures to enable sky subtraction and to avoid hot/dead pixels. Relative astrometry was obtained through a Gaussian PSF fit, consistent with methods used in previous Keck observations of TNBs \citep[e.g.][]{grundy2011five}. Satellite astrometry derived from the February 2023 stellar occultation and the averaged positions from the March 2021 and June 2023 events were added to the astrometry database. 

The combined astrometric data, obtained over 20 years (Table \ref{tab:binary_astrometry}), provides a powerful dataset to calculate the Huya system's mutual orbit. The orbit fitting was completed using MultiMoon, a Markov Chain Monte Carlo (MCMC) orbit fitting approach described in \citet{ragozzine2024beyond1} and \citet{proudfoot2024beyond3, proudfoot2024beyond2}. See \citet{hogg2018data} for a primer on MCMC methods. The orbit fit was run under the assumption of Keplerian motion (i.e., no orbital precession); we will review this assumption later. To find the global best fit, dozens of orbit fits were completed, testing initial walker positions across the 7-dimensional orbit parameter space (system mass, semi-major axis, eccentricity, and four-orbit orientation angles). 

\begin{table}[!htb]
    \centering
    \caption{Huya's satellite relative right ascension ($\alpha$) and declination ($\delta$) ordered by date. The astrometry from the March 2021 and June 2023 stellar occultation events consider the average of both astrometric solutions from Table \ref{tab:astrometry_results2}.}
    \label{tab:binary_astrometry}
    \resizebox{14cm}{!}{
    \begin{tabular}{ccccccc}
    \hline \hline
Julian Date & Date & Telescope/Instrument & $\Delta \alpha \cos{\delta}$ & $\sigma_{\Delta \alpha \cos{\delta}}$ & $\Delta \delta$  & $\sigma_{\Delta \delta}$ \\
            &      &                      & ($''$) & ($''$)   & ($''$) & ($''$)      \\
\hline
2452456.328 & 2002-06-30 & HST/STIS & -0.08072 & 0.00256 & -0.01525 & 0.00213 \\
2452457.131 & 2002-07-01 & HST/STIS & -0.03174 & 0.00338 & 0.06966 & 0.00245 \\
2456053.542 & 2012-05-06 & HST/WFC3 & -0.08701 & 0.00174 & 0.02700 & 0.00136 \\
2459301.559 & 2021-03-28 & Occultation & -0.04146 & 0.00814 & 0.08291 & 0.00521 \\
2459393.953 & 2021-06-28 & Keck/NIRC2 & -0.05270 & 0.00300 & -0.03062 & 0.00300 \\
2459394.957 & 2021-06-29 & Keck/NIRC2 & -0.04217 & 0.00300 & 0.07594 & 0.00300 \\
2459992.988 & 2023-02-17 & Occultation & -0.05503 & 0.00713 & -0.00518 & 0.00930 \\
2460119.540 & 2023-06-24 & Occultation & 0.06546 & 0.00300 & -0.04440 & 0.00500 \\ 
\hline
\end{tabular} 
    }
\end{table}

Once a preferred orbit solution was found in initial exploratory fits, a long orbital fit was executed, starting near the center of the preferred solution. The MCMC orbit fit used 1000 walkers--simultaneously runs of Markov chains--and started with a 10000 step burn-in phase, after which the Markov chains were discarded. After the burn-in, poorly performing walkers were removed and replaced with random linear combinations of highly-performing walkers, after which another 1000-step burn-in phase was run and discarded. The final ensemble of walkers ran for 25000 steps \citep[see][for more details on the MCMC fitting procedures]{foreman2013emcee,ragozzine2024beyond1, proudfoot2024beyond2}. Convergence of the fit was assessed by visual inspection of walker trace plots, marginal posteriors, and joint posteriors (see Appendix \ref{sec:appendixe}). In total, this single fit tested over 30 million sets of orbit parameters against the data. Including preliminary exploration runs, $\sim$500 million tests of possible orbit parameters were performed. 

\begin{figure}[!htb]
    \centering
    \includegraphics[width=0.62\linewidth]{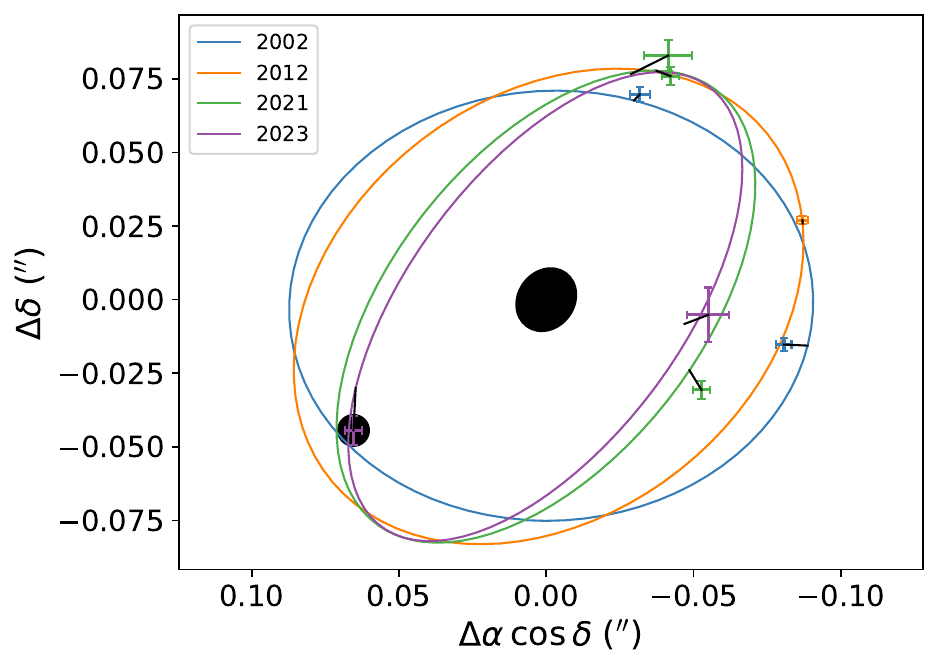}
    \caption{A comparison of the Keplerian orbit solution and the observational data. The black ellipses are approximately to scale and show the shape of Huya and its satellite during the 2023-06-24 occultation. Colored crosses show relative astrometry at the various epochs of observation. Black lines connect the observations to the position predicted by the Keplerian model. Colored ellipses show the orbit during the first observation in that calendar year. The differences in apparent orbit over time show the changing opening angle of the satellite's orbit (see Sect. \ref{sec:discussion}). 
    }
    \label{fig:orbit}
\end{figure}

Despite the large volume of orbit parameters tested in this work, the best-fit Keplerian orbit (shown in Table \ref{tab:binary_orbit} and Figure \ref{fig:orbit}) had $\chi^2 \sim 36$ with 9 degrees of freedom, giving $\chi^2_{pdf} \sim 4$. Although the fit presents statistically poor quality, the typical residuals on the observations are relatively small, with RMS residuals of 9 mas, comparable to the size of a pixel on Keck or 25\% of an HST pixel. The chance that a true Keplerian orbit would produce as bad (or worse) of a fit is $\sim5 \times 10^{-5}$ or 1-in-20,000. The poor quality is likely the result of one of two possible issues: i) low-quality data contaminating the relative astrometry dataset, or ii) non-Keplerian motion causing a poor-quality Keplerian orbit fit. The eight astrometry measurements are of high quality, and the data processing pipeline has been validated over more than a decade of use \citep[e.g.,][]{grundy200842355, grundy2009mutual}. Still, data contamination is always possible, and we cannot discard it as a possibility. The Huya system is of particular concern because the maximum separation between both components is $\sim0.1''$, which approaches the resolution limits of HST and Keck. Compared with HST WFC3, JWST's NIRCam has a slightly higher resolution pixel scale (0.03 $''$/px compared to 0.04 $''$/px) and much better PSF FWHM (0.029$''$ compared to 0.067$''$), providing a platform to test the data contamination hypothesis.

\begin{table}[!htb]
    \centering
    \caption{Huya's satellite orbit fitted and derived parameters. All orbital angles relate to the J2000 ecliptic frame on JD 2452400 (2002-05-05 12:00 UTC), except for the derived orbit pole ($\alpha$, $\delta$) which are given in the J2000 equatorial frame. }
    \label{tab:binary_orbit}
    \resizebox{13.9cm}{!}{
    \begin{tabular}{lccc}
    \hline \hline
Fitted parameters & & Posterior Distribution & Best Fit \\
\hline
\qquad System Mass ($10^{18}$ kg) & $M_{sys}$ & $45.2^{+1.6}_{-1.5}$ & $44.9$ \\
\qquad Semi-major axis (km) & $a$ & $1898^{+22}_{-21}$ & $1895$ \\
\qquad Eccentricity & $e$ & $0.036^{+0.017}_{-0.015}$ & $0.034$ \\
\qquad Inclination ($^{\circ}$) & $i$ & $65.8^{+1.9}_{-1.9}$ & $65.8$ \\
\qquad Argument of periapse ($^{\circ}$) & $\omega$ & $101^{+17}_{-24}$ & $100$ \\
\qquad Longitude of the ascending node ($^{\circ}$) & $\Omega$ & $122.9^{+1.7}_{-1.6}$ & $122.9$ \\
\qquad Mean anomaly at epoch ($^{\circ}$) & $\mathcal{M}$ & $147^{+23}_{-17}$ & $147$ \\
\hline
Derived parameters & & & \\
\hline
\qquad Orbit period (days) & $P_{orb}$ & $3.46293^{+0.00001}_{-0.00001}$ & $3.46293$ \\
\qquad Orbit Pole Right Ascension ($^{\circ}$) & $\alpha$ & $20.8^{+1.9}_{-1.9}$ & $20.8$ \\
\qquad Orbit Pole Declination ($^{\circ}$) & $\delta$ & $34.9^{+1.9}_{-1.9}$ & $34.9$ \\
\hline

\hline
\end{tabular} 
    }
\end{table}

Non-Keplerian effects could cause a low-quality orbit fit like the one we present here. Given the system's tight mutual orbit (maximum separation $\sim$2000 km) and short orbit period ($\sim$3.5 days), any putative precession would be easily detectable over the 20-year observational baseline. Using the 2-dimensional projected shape of Huya obtained with the \textit{2023 Restrict} approach described in Section \ref{sec:data_analysis} (giving a $J_2 \approx 0.04$) and the analytical formula for precession in TNBs \citep{proudfoot2024beyond2}, the orbital precession period of Huya's satellite would be $\lesssim5$ years, implying that the satellite's orbit may have precessed a few times since its discovery. Hence, with even a small eccentricity or inclination (with respect to Huya's equator), substantial deviations from Keplerian motion are expected. Therefore, given the small eccentricity detected in the satellite's orbit, the non-Keplerian effects are a good explanation for the poor quality fit presented above. 

A brief test of this hypothesis was performed using a non-Keplerian orbit fit to the observational data. An orbit fit with $\chi^2 \sim 14$ ($\chi_{pdf}^2 \sim 3$) was obtained, which is more than the expected improvement from the additional degrees of freedom. The random chance that this fit quality (or worse) could be achieved by a true non-Keplerian orbit is $\sim$0.02, or a 1-in-50 chance, still somewhat worse than desired. Therefore, either the fit did not fully converge or the errors in our observations may be somewhat underestimated. Although likely not the best global fit, the non-Keplerian fit has a similar mass, semi-major axis, and orbital period compared to the Keplerian fit presented above. The orbital solution has reasonable non-Keplerian orbit parameters with an estimated apsidal (nodal) precession period of $\sim1-1.5$ ($\sim2-3$) years. This drastic improvement in quality shows that non-Keplerian effects are likely to be responsible for the poor quality fit while confirming that the Keplerian orbit fit still captures essential information about the system (e.g., mass, semi-major axis, period). Small changes in the mutual orbit properties are expected to occur after non-Keplerian analysis, but they are unlikely to change the overarching findings. Due to the complexity involved in performing complete non-Keplerian orbit fits—particularly in the context of rapid precession—we will defer this fitting problem to future research.

\section{Discussion and conclusions}
\label{sec:discussion}

In this work, we present the three stellar occultation events by the Huya system, the binary with the second shortest mutual orbit among the known Trans-Neptunian Binaries (TNBs), after Lempo-Hiisi. The June 2023 event is the second known multi-chord stellar occultation by Huya, and the three limb solutions here obtained (shown in Table \ref{tab:fitted parameters}) agree at the $ 1\sigma$ level with the Huya profile published by \cite{SantosSanz2022}, except for the position angle interval obtained from the \textit{2023 Free} approach. Since the 2019 stellar occultation event, Huya only moved $\approx$ 1.67 $\%$ in its orbit around the Sun, changing the aspect angle by only a few degrees. Therefore, the observed discrepancy of approximately 20$^\circ$ in the ellipse position angle between the 2019 and 2023 stellar occultation events can only be attributed to a putative triaxial shape. However, the low amplitude of the published rotational light curve \citep{SantosSanz2022} does not support such a shape for Huya. Assuming that the satellite orbits at Huya's equatorial plane, the shallow rotational light curve reported by \cite{SantosSanz2022} cannot be explained by a pole-on observational orientation. In this context, our preferred limb solution comes from the \textit{2023 Restrict} approach (Fig. \ref{fig:Huya_limb_fittings}b). 

On the other hand, the roughly unchanged projected area since 2019, along with the small amplitude of the rotational light curve, suggests an oblate or Maclaurin shape for Huya. Therefore, we determined Huya's global density using two distinct methods: i) the mass and volume of the binary system to obtain the system density of $\rho_1 = 1073\ \pm$ 66 kg~m$^{-3}$ (which assumes a spherical satellite and the same density for both components) and ii) Huya's rotational period and the Chandrasekhars’ formalism \citep{Chandrasekhar1969} to obtain the primary density of $\rho_2$ = 768 $\pm$ 42 kg~m$^{-3}$. Assuming that Huya has a Maclaurin tridimensional shape and the density indicated by the second solution, then a satellite with a diameter of approximately 200 km would need to have a density of $\approx$ 3500 kg~m$^{-3}$ to match the total system density obtained from the first method. This would imply that the satellite is the densest trans-Neptunian object (TNO) ever identified, a proposition that appears highly unlikely. An alternative approach is to assume that Huya and its satellite share the same density, so Huya alone has a density of 1073 kg~m$^{-3}$, as derived from the first solution. Comparing this with the second solution, the discrepancy suggests that Huya is likely not conforming to the Maclaurin equilibrium shape. This is plausible given that Huya's diameter is near the 450 km limit for which hydrostatic equilibrium is expected \citep{Tancredi2008}. Therefore, based on the derived profiles and the published rotational light curve, an oblate figure with a density of $\rho_1 = 1073$ kg~m$^{-3}$ is our preferred solution for Huya. 

In addition to Huya, we also present three single-chord detections of Huya's satellite from the 2021 and 2023 stellar occultation events. The best limb measurement from June 2023 puts a lower limit for the diameter of D = 165 km, assuming a spherical satellite.  The satellite's absolute magnitude of $H_V$ = 6.68 $\pm$ 0.18 mags was calculated from the flux difference between the system's and Huya's absolute magnitudes of $H_V$ = 5.04 $\pm$ 0.03 mag and $H_V$ = 5.31 $\pm$ 0.03 mags, respectively \citep{SantosSanz2022}. Therefore, an upper limit for the satellite geometric albedo can be obtained, $p_V$ = 0.15. Higher than Huya geometric albedo ($p_V$ = 0.079 $\pm$ 0.004), but still fully dependent on the assumptions we made about the satellite size and shape. However, if such high albedo is confirmed, it would be the second example of a bright satellite orbiting a TNO after Hi'iaka \citep{Estela2022EPSC}.

Single-chord stellar occultations can provide valuable astrometry for improving an object's orbit solution \citep{Rommel2020}, moreover when this object is a satellite with no prior orbit determination. Using the relative astrometry obtained from the stellar occultations, along with the astrometry obtained through HST and Keck images, we obtained an orbit for Huya's satellite. As the Keplerian approach presents a low-quality orbit (Table \ref{tab:binary_orbit}), we also tested the hypothesis of a non-Keplerian orbit, which results in similar parameters compared to the Keplerian fit. A full non-Keplerian orbit fit in the fast-precession domain is complex and will be the topic of future work.

According to our results, the satellite orbit opening angle is slowly decreasing over the observations' time frame, as seen in Figure \ref{fig:orbit}. Therefore, given our derived Keplerian orbit and the derived sizes of Huya and its satellite, the binary's mutual event season is expected to begin in approximately 2033. At first, they will be short grazing events but eventually will grow in duration and depth in the following years. The mutual event season will peak in $\sim$2039, with events that last $\sim$5 hours where the total system will dim by $\sim$0.25 mag (based on the absolute magnitudes of Huya and the combined system provided above). Since the system is relatively bright ($V = 19-20$ mag) and events reoccur twice every 3.46 days for several years, observing these events will be relatively accessible to even 0.5-1 meter telescopes. In this context, before the beginning of the mutual event season, high-resolution observations (from HST, JWST, or possibly Keck) should be taken to provide a precise schedule for the upcoming events. Mutual events can provide information about various system properties, including size, shape, albedo, superficial albedo variegation, and mutual orbit properties. Given the unique chance of observing a tight binary TNO during mutual events, the community should consider a long-term preparedness plan.

\begin{figure}[!htb]
    \centering
    \includegraphics[width=\linewidth]{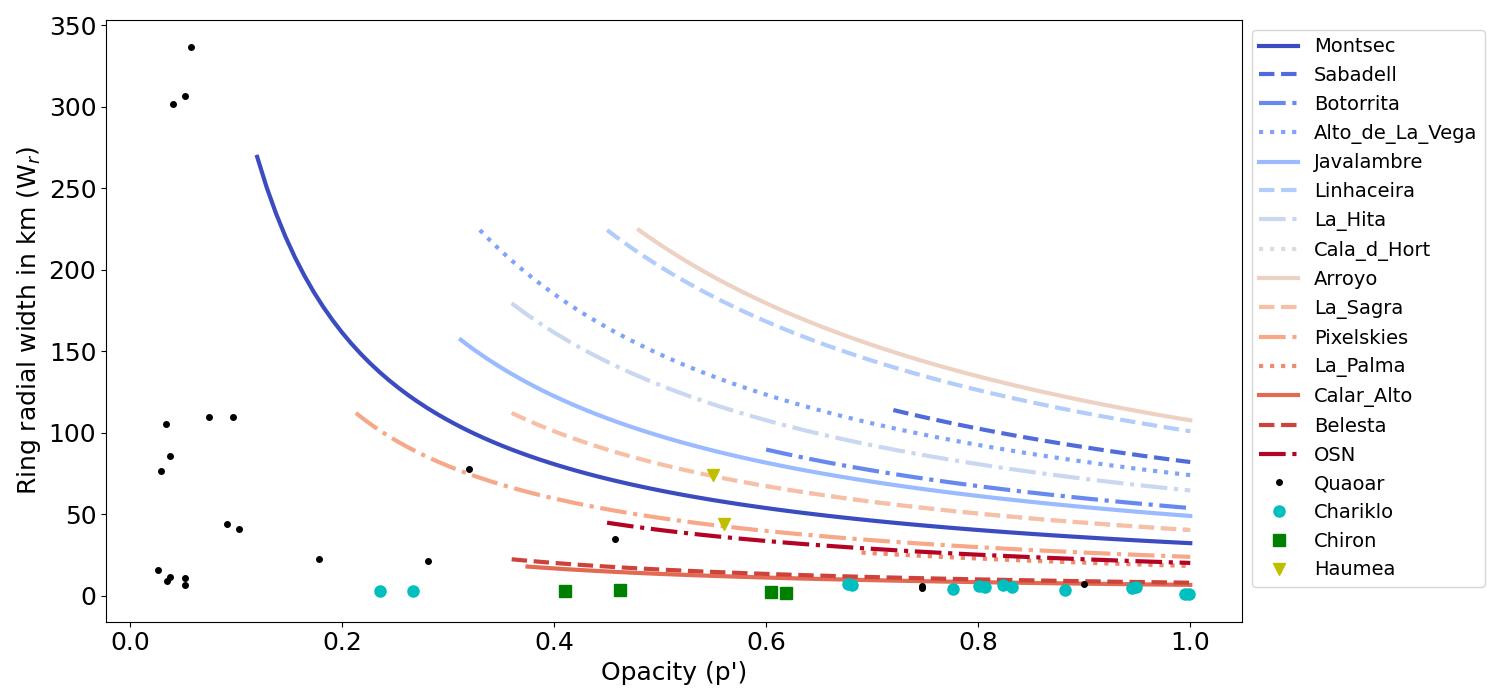}
    \caption{Ring radial width as a function of the ring apparent opacity ($p'$), considering the data dispersion and exposure time of each data set (see text). The data points represent the known rings around small bodies.}
    \label{fig:ring_features}
\end{figure}

\begin{figure}[!htb]
    \centering
    \includegraphics[width=\linewidth]{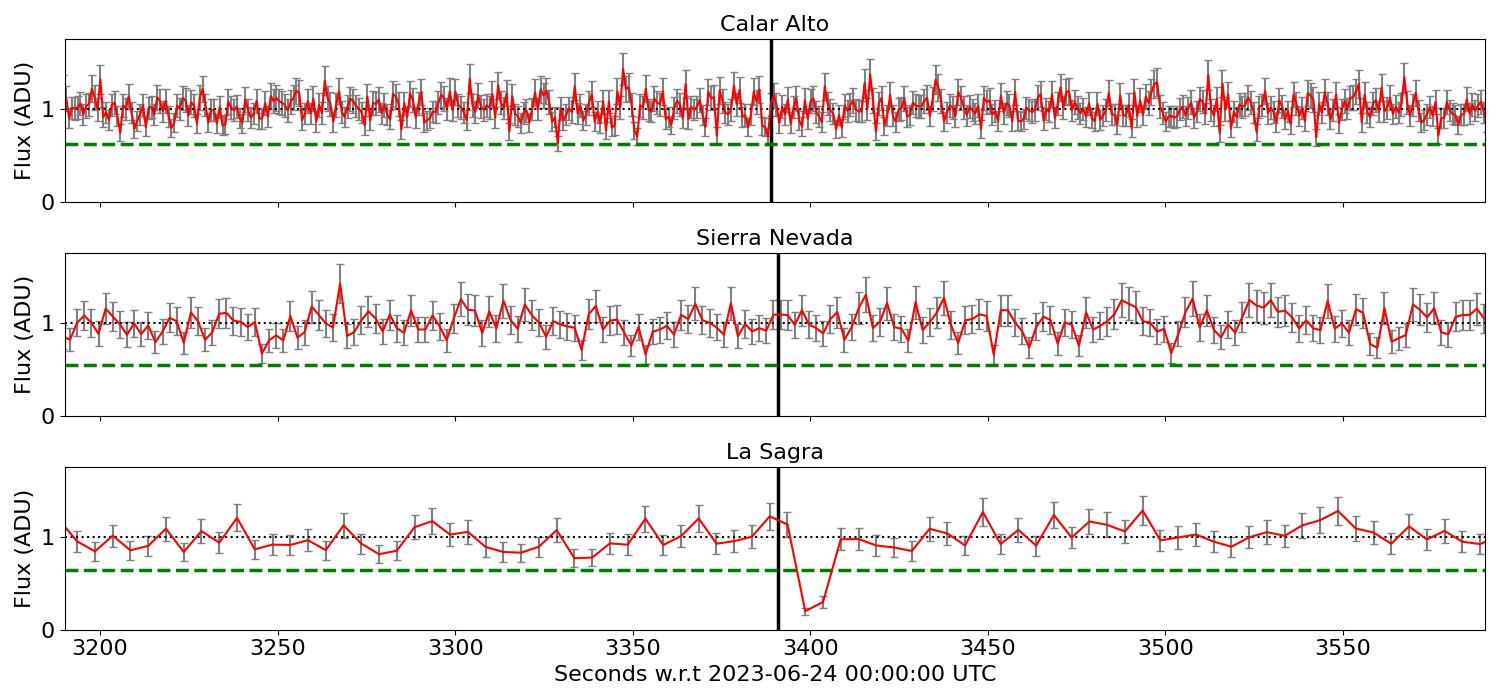}
    \caption{Target star flux ratio (red) with uncertainties (gray) as a function of time for the three datasets that place the best constraints on the presence of rings around Huya (see text). The vertical black line presents the prediction instant for the Huya occultation at each site and the green dashed line shows the $3\sigma$ detection level on each light curve.}
    \label{fig:ring_constraining_LC}
\end{figure}

 A lower limit for a putative ring system surrounding Huya also was determined by following the same procedures described in \cite{SantosSanz2022} and \cite{Estela2023}. The ring apparent width can be calculated using equation \ref{eq:ring_width},
 \begin{equation}
     W'=\frac{3\sigma v T_{exp}}{p'}
     \label{eq:ring_width}
 \end{equation}
where $\sigma$ corresponds to the light curve dispersion, $v$ is Huya's apparent sky velocity at the moment of the occultation (km/s), $T_{exp}$ is the exposure time (s) used for each data set, and $p'$ is the putative ring apparent opacity. An exploration between opacities of $3\sigma<p'<1$ was made, and the most accurate data sets are shown in Figure \ref{fig:ring_features}. The most stringent constraint for the presence of rings is provided by the Calar Alto negative dataset, which probed the surroundings for structures as narrow as 8 km for $p'$ = 1 to broad 18 km rings for $p'$ = 0.45. The dead time in these data corresponds to an uncertainty of 1.4 km in the sky. Therefore, to be above the 3$\sigma$ level in the Calar Alto data, the ring-like structure should have apparent opacity greater than 45$\%$ and a width greater than 9.4 km. The La Palma light curve has a dead time of 0.05 seconds but is noisier and would only detect opaque rings ($p'>0.69$) with apparent widths greater than 18 km. Lastly, the La Sagra light curve does not have dead times and could probe for ring-like widths from 40 to 112 km ($p'=1.0$ to $0.36$). The Belesta and Montsec data sets, despite seeming promising in Figure \ref{fig:ring_features}, have dead times greater than 2 seconds, which means an uncertainty greater than 45 km and could not place meaningful constraints on the presence of rings surrounding Huya. Therefore, considering all known small body ring-like structure apparent opacities and radial widths ($p'$, $W_r$) recovered from the literature, the data sets we obtained here would not be able to detect most of them. The only ring system that, if present in Huya,  would be detected in the Calar Alto, Sierra Nevada (OSN), and La Sagra light curves is a Haumea-like ring. However, no evidence of flux drops above the $3\sigma$ confidence level appears in a range of $\approx$ 9000 km centered in the main body predicted location in the light curve, other than the detection of Huya in the La Sagra light curve (see Fig. \ref{fig:ring_constraining_LC}).

\pagebreak

\section*{Acknowledgments}
SORA team members R. C. Boufleur, G. Margotti, and M. Banda-Huarca, for their support during the data analysis procedures. To the following observers for their participation in the observations: F. Campos-García, N. Graciá, A. Popowicz, V. Pelenjow, F. Ursache.
This study was partially financed by the  National Institute of Science and Technology of the e-Universe project (INCT do e-Universo, CNPq grant 465376/2014-2).
The staff of ESO La Silla, Pico dos Dias Observatory, and Sierra Nevada Observatory for their support during astrometric runs on those telescopes. This research made use of \textsc{ccdproc}, an \textsc{Astropy} package for image reduction \citep{Craig2023}. 
We also acknowledge financial support from the Severo Ochoa grant CEX2021-001131-S funded by MCIN/AEI/10.13039/501100011033 and by the Spanish projects PID2020-112789GB-I00 from AEI and Proyecto de Excelencia de la Junta de Andalucía PY20-01309. 
F.L.R. acknowledges CNPq, Brazil grant 103096/2023-0, Florida Space Institute's Space Research Initiative, and the University of Central Florida's Preeminent Postdoctoral Program (P3). B. C. N. Proudfoot acknowledges the University of Central Florida's Preeminent Postdoctoral Program (P3).
M.A. acknowledges CNPq, Brazil grants 427700/2018-3, 310683/2017-3, and 473002/2013-2. 
F.B-R. acknowledges CNPq grant 316604/2023-2.
P.S-S. acknowledges financial support from the Spanish I+D+i project PID2022-139555NB-I00 (TNO-JWST) funded by MCIN/AEI/ 10.13039/501100011033. 
T.S-R acknowledges funding from Ministerio de Ciencia e Innovación (Spanish Government), PGC2021, PID2021-125883NB-C21. This work was (partially) supported by the Spanish MICIN/AEI/10.13039/501100011033 and by ``ERDF A way of making Europe" by the “European Union” through grant PID2021-122842OB-C21, and the Institute of Cosmos Sciences University of Barcelona (ICCUB, Unidad de Excelencia `María de Maeztu’) through grant CEX2019-000918-M. The Joan Oró Telescope (TJO) of the Montsec Astronomical Observatory (OAdM) is owned by the Catalan Government and operated by the Institute for Space Studies of Catalonia (IEEC).
The work of K.H. was supported by the project RVO:67985815.
A.J.C-T acknowledges support from the Spanish Ministry projects PID2020-118491GB-I00 and PID2023-151905OB-I00 and Junta de Andaluc\'ia grant P20$\_$010168.
V. N. acknowledges financial support from the Bando Ricerca Fondamentale INAF 2023, Data Analysis Grant: “Characterization of transiting exoplanets by exploiting the unique synergy between TASTE and TESS”.
C.L.P is thankful for the support of the Coordenação de Aperfeiçoamento de Pessoal
de Nível Superior - Brasil (CAPES) and FAPERJ/DSC-10 (E26/204.141/2022).
W.G. is grateful for support from a NASA Keck PI Data Award, administered by the NASA Exoplanet Science Institute. Data presented herein were obtained at the W. M. Keck Observatory from telescope time allocated to NASA through the agency's scientific partnership with the California Institute of Technology and the University of California. The Observatory was made possible by the generous financial support of the W. M. Keck Foundation. The authors wish to recognize and acknowledge the very significant cultural role and reverence that the summit of Maunakea has always had within the indigenous Hawaiian community. We are most fortunate to have the opportunity to conduct observations from this mountain. 
This research has made use of the Keck Observatory Archive (KOA), which is operated by the W. M. Keck Observatory and the NASA Exoplanet Science Institute (NExScI), under contract with the National Aeronautics and Space Administration. 
This work makes use of observations collected at the Asiago Schmidt 67/92 cm telescope (Asiago, Italy) of the INAF -- Osservatorio Astronomico di Padova. 
Partially based on observations made with the Tx40 telescope at the Observatorio Astrofísico de Javalambre in Teruel, a Spanish Infraestructura Cientifico-Técnica Singular (ICTS) owned, managed, and operated by the Centro de Estudios de Física del Cosmos de Aragón (CEFCA). Tx40 is funded with the Fondos de Inversiones de Teruel (FITE). This research is based on observations made with the NASA/ESA Hubble Space Telescope obtained from the Space Telescope Science Institute, which is operated by the Association of Universities for Research in Astronomy, Inc., under NASA contract NAS 5–26555. These observations are associated with programs 9110 and 12468 .

\newpage
\appendix
\label{appendix:A}
\section{Observational circumstances}
\label{sec:appendixd}

\begin{table}[!h]
\resizebox{\linewidth}{!}{
\begin{tabular}{|l|l|l|l|l|l|}
\hline
\multicolumn{1}{|c|}{\textbf{\begin{tabular}[c]{@{}c@{}}Observatory,\\ nearest city,\\ country\end{tabular}}} &
  \multicolumn{1}{c|}{\textbf{\begin{tabular}[c]{@{}c@{}}Latitude ($^\circ$),\\ longitude ($^\circ$),\\ height (m)\end{tabular}}} &
  \multicolumn{1}{c|}{\textbf{\begin{tabular}[c]{@{}c@{}}Telescope,\\ aperture (m),\\ filter\end{tabular}}} &
  \multicolumn{1}{c|}{\textbf{\begin{tabular}[c]{@{}c@{}}Time source,\\ instrument,\\ written time\end{tabular}}} &
  \multicolumn{1}{c|}{\textbf{\begin{tabular}[c]{@{}c@{}}Exposure (s),\\ cycle (s),\\ offset (s)\end{tabular}}} &
  \multicolumn{1}{c|}{\textbf{Observers}} \\ \hline
  \multicolumn{6}{|c|}{2021-03-28}\\ \hline
\begin{tabular}[c]{@{}l@{}}-\\Ond\v{r}ejov, \\ Czech Republic\end{tabular} &
  \begin{tabular}[c]{@{}l@{}}49.910560,\\ 14.78364000,\\ 528.0\end{tabular} &
  \begin{tabular}[c]{@{}l@{}}-\\ 0.65,\\ clear\end{tabular} &
  \begin{tabular}[c]{@{}l@{}}PC +  NTP,\\ Moravian G2-3200,\\ Start of Exposure\end{tabular} &
  \begin{tabular}[c]{@{}l@{}}8.0,\\ 9.5,\\ -\end{tabular} &
  \begin{tabular}[c]{@{}l@{}}H. Kučáková,\\K. Hornoch  \\ \end{tabular} \\ \hline
  \multicolumn{6}{|c|}{2023-02-17}\\ \hline
\begin{tabular}[c]{@{}l@{}}Penrose\\ Colorado,\\ Unites States of America\end{tabular} &
  \begin{tabular}[c]{@{}l@{}}38.4683207,\\ -104.9899452,\\ 1660.0\end{tabular} &
  \begin{tabular}[c]{@{}l@{}}SCT,\\ 0.2794,\\ None\end{tabular} &
  \begin{tabular}[c]{@{}l@{}}GPS,\\ QHY174M-GPS,\\ Start of Exposure\end{tabular} &
  \begin{tabular}[c]{@{}l@{}}5.0,\\ 5.18,\\ - \end{tabular} &
  \begin{tabular}[c]{@{}l@{}} V. Nikitin \end{tabular} \\ \hline
\multicolumn{6}{|c|}{2023-06-24}\\ \hline
  \multicolumn{1}{|l|}{\begin{tabular}[c]{@{}l@{}}Montsec, \\ Lleida,\\ Spain\end{tabular}} &
  \multicolumn{1}{l|}{\begin{tabular}[c]{@{}l@{}}42.051655,\\ 0.72965,\\ 1564.582\end{tabular}} &
  \multicolumn{1}{l|}{\begin{tabular}[c]{@{}l@{}}TJO,\\ 0.8,\\ R (692.392 $\pm$ 139.986 nm)\end{tabular}} &
  \multicolumn{1}{l|}{\begin{tabular}[c]{@{}l@{}}GPS,\\ CCD42-40,\\ Start of Exposure\end{tabular}} &
  \multicolumn{1}{l|}{\begin{tabular}[c]{@{}l@{}}12.0,\\ 14.69,\\ -\end{tabular}} &
  \multicolumn{1}{l|}{\begin{tabular}[c]{@{}l@{}}~\\ T. Santana-Ros\\ ~\end{tabular}} \\ \hline
   \multicolumn{1}{|l|}{\begin{tabular}[c]{@{}l@{}}Botorrita, \\ Zaragoza,\\ Spain\end{tabular}} &
  \multicolumn{1}{l|}{\begin{tabular}[c]{@{}l@{}}41.497375,\\ -1.020867,\\ 403.0\end{tabular}} &
  \multicolumn{1}{l|}{\begin{tabular}[c]{@{}l@{}}Stargate,\\ 0.50,\\ Clear\end{tabular}} &
  \multicolumn{1}{l|}{\begin{tabular}[c]{@{}l@{}}GPS,\\ QHY174,\\ Start of Exposure\end{tabular}} &
  \multicolumn{1}{l|}{\begin{tabular}[c]{@{}l@{}}4.0,\\ 4.32,\\ -\end{tabular}} &
  \multicolumn{1}{l|}{\begin{tabular}[c]{@{}l@{}}O. Canales, D. Lafuente, \\ S. Calavia, F. Campos\end{tabular}} \\ \hline
    \multicolumn{1}{|l|}{\begin{tabular}[c]{@{}l@{}}Sabadell,\\ Catalonia,\\ Spain\end{tabular}} &
  \multicolumn{1}{l|}{\begin{tabular}[c]{@{}l@{}}41.550043,\\ 2.09013,\\ 224.0\end{tabular}} &
  \multicolumn{1}{l|}{\begin{tabular}[c]{@{}l@{}}Newtonian,\\ 0.50,\\ Empty\end{tabular}} &
  \multicolumn{1}{l|}{\begin{tabular}[c]{@{}l@{}}GPS, \\ Watec 910HX/RC,\\ Middle of Exposure\end{tabular}} &
  \multicolumn{1}{l|}{\begin{tabular}[c]{@{}l@{}}5.08,\\ 5.18,\\ -2.54\end{tabular}} &
  \multicolumn{1}{l|}{\begin{tabular}[c]{@{}l@{}}C. Perelló, \\ A. Selva\end{tabular}} \\ \hline
\multicolumn{1}{|l|}{\begin{tabular}[c]{@{}l@{}}Alto de La Vega,\\ Vega del Codorno,\\ Spain\end{tabular}} &
  \multicolumn{1}{l|}{\begin{tabular}[c]{@{}l@{}}40.4172959742339,\\ -1.9106369708525,\\ 1533.0\end{tabular}} &
  \multicolumn{1}{l|}{\begin{tabular}[c]{@{}l@{}}RCT,\\ 0.3556,\\ Luminance\end{tabular}} &
  \multicolumn{1}{l|}{\begin{tabular}[c]{@{}l@{}}PC+NTP,\\ ASI 6200MM Pro,\\ Start of Exposure\end{tabular}} &
  \multicolumn{1}{l|}{\begin{tabular}[c]{@{}l@{}}10.0,\\ 10.37,\\ -\end{tabular}} &
  \multicolumn{1}{l|}{\begin{tabular}[c]{@{}l@{}}E. García-Navarro,\\ J. E. Donate-Lucas,\\ L. Izquierdo-Carrión\end{tabular}} \\ \hline
\multicolumn{1}{|l|}{\begin{tabular}[c]{@{}l@{}}Observatorio Astrofísico \\ de Javalambre, \\ Spain\end{tabular}} &
  \multicolumn{1}{l|}{\begin{tabular}[c]{@{}l@{}}40.0418,\\ -1.0163,\\ 1957.0\end{tabular}} &
  \multicolumn{1}{l|}{\begin{tabular}[c]{@{}l@{}}Cassegrain,\\ 0.40,\\ Clear\end{tabular}} &
  \multicolumn{1}{l|}{\begin{tabular}[c]{@{}l@{}}PC+NTP,\\ ProLine PL4720,\\ Start of Exposure\end{tabular}} &
  \multicolumn{1}{l|}{\begin{tabular}[c]{@{}l@{}}7.0,\\ 7.78,\\ -\end{tabular}} &
  \multicolumn{1}{l|}{\begin{tabular}[c]{@{}l@{}}R. Iglesias-Marzoa, \\ E. Lacruz\end{tabular}} \\ \hline
\multicolumn{1}{|l|}{\begin{tabular}[c]{@{}l@{}}Linhaceira, \\ Portugal\end{tabular}} &
  \multicolumn{1}{l|}{\begin{tabular}[c]{@{}l@{}}39.522688,\\ -8.3838,\\ 90.0\end{tabular}} &
  \multicolumn{1}{l|}{\begin{tabular}[c]{@{}l@{}}SCT,\\ 0.355,\\ Clear\end{tabular}} &
  \multicolumn{1}{l|}{\begin{tabular}[c]{@{}l@{}}PC+DCF77,\\ SBIG ST7-XME,\\ Start of Exposure\end{tabular}} &
  \multicolumn{1}{l|}{\begin{tabular}[c]{@{}l@{}}10.0,\\ 12.96,\\ -\end{tabular}} &
  \multicolumn{1}{l|}{\begin{tabular}[c]{@{}l@{}}~\\ R. Gonçalves\\ ~\end{tabular}} \\ \hline
\multicolumn{1}{|l|}{\begin{tabular}[c]{@{}l@{}}La Hita, \\ Spain\end{tabular}} &
  \multicolumn{1}{l|}{\begin{tabular}[c]{@{}l@{}}39.568,\\ -3.1833,\\ 770.0\end{tabular}} &
  \multicolumn{1}{l|}{\begin{tabular}[c]{@{}l@{}}Newtonian,\\ 0.77\\ Empty\end{tabular}} &
  \multicolumn{1}{l|}{\begin{tabular}[c]{@{}l@{}}PC+NTP,\\ SBIG STL11000 - SOE,\\ Start of Exposure\end{tabular}} &
  \multicolumn{1}{l|}{\begin{tabular}[c]{@{}l@{}}8.0,\\ 11.21,\\ -\end{tabular}} &
  \multicolumn{1}{l|}{\begin{tabular}[c]{@{}l@{}}N. Morales, \\ F. Organero, \\ L. Hernández\end{tabular}} \\ \hline
\multicolumn{1}{|l|}{\begin{tabular}[c]{@{}l@{}}Cala d' Hort,\\ Baleares,\\ Spain\end{tabular}} &
  \multicolumn{1}{l|}{\begin{tabular}[c]{@{}l@{}}38.891102,\\ 1.2408,\\ 160.0\end{tabular}} &
  \multicolumn{1}{l|}{\begin{tabular}[c]{@{}l@{}}TCH,\\ 0.51,\\ Luminance\end{tabular}} &
  \multicolumn{1}{l|}{\begin{tabular}[c]{@{}l@{}}PC+NTP,\\ ASI 6200MM Pro mono,\\ Start of Exposure\end{tabular}} &
  \multicolumn{1}{l|}{\begin{tabular}[c]{@{}l@{}}3.0,\\ 3.42,\\ -\end{tabular}} &
  \multicolumn{1}{l|}{\begin{tabular}[c]{@{}l@{}}I. de la Cueva, \\ M. Moreno\end{tabular}} \\ \hline
\multicolumn{1}{|l|}{\begin{tabular}[c]{@{}l@{}}Arroyo,\\ Murcia,\\ Spain\end{tabular}} &
  \multicolumn{1}{l|}{\begin{tabular}[c]{@{}l@{}}38.0968434847573,\\ -1.675721238961636,\\ 468.7075500488281\end{tabular}} &
  \multicolumn{1}{l|}{\begin{tabular}[c]{@{}l@{}}LX200,\\ 0.30,\\ Empty\end{tabular}} &
  \multicolumn{1}{l|}{\begin{tabular}[c]{@{}l@{}}PC+NTP,\\ ST-1001E,\\ Start of Exposure\end{tabular}} &
  \multicolumn{1}{l|}{\begin{tabular}[c]{@{}l@{}}10.0,\\ 13.0,\\ -\end{tabular}} &
  \multicolumn{1}{l|}{\begin{tabular}[c]{@{}l@{}}J. Reyes, \\ S. Pastor\end{tabular}} \\ \hline
\multicolumn{1}{|l|}{\begin{tabular}[c]{@{}l@{}}La Sagra, \\ Spain\end{tabular}} &
  \multicolumn{1}{l|}{\begin{tabular}[c]{@{}l@{}}37.981,\\ -2.564,\\ 1530.0\end{tabular}} &
  \multicolumn{1}{l|}{\begin{tabular}[c]{@{}l@{}}Tetrascopio,\\ 0.356,\\ Empty\end{tabular}} &
  \multicolumn{1}{l|}{\begin{tabular}[c]{@{}l@{}}GPS,\\ QHY174M,\\ Start of Exposure\end{tabular}} &
  \multicolumn{1}{l|}{\begin{tabular}[c]{@{}l@{}}5.0,\\ 5.0,\\ -\end{tabular}} &
  \multicolumn{1}{l|}{\begin{tabular}[c]{@{}l@{}}~\\ N. Morales\\ ~\end{tabular}} \\ \hline
\multicolumn{1}{|l|}{\begin{tabular}[c]{@{}l@{}}PixelSkies, \\ Granada, \\ Spain\end{tabular}} &
  \multicolumn{1}{l|}{\begin{tabular}[c]{@{}l@{}}37.739722,\\ -2.643889,\\ 805.0\end{tabular}} &
  \multicolumn{1}{l|}{\begin{tabular}[c]{@{}l@{}}TAGRA,\\ 0.508,\\ Clear\end{tabular}} &
  \multicolumn{1}{l|}{\begin{tabular}[c]{@{}l@{}}PC+NTP,\\ ASI 1600MM Pro Mono,\\ Start of Exposure\end{tabular}} &
  \multicolumn{1}{l|}{\begin{tabular}[c]{@{}l@{}}5.0,\\ 7.2,\\ +2.5\end{tabular}} &
  \multicolumn{1}{l|}{\begin{tabular}[c]{@{}l@{}}B. Staels, R. Goossens, \\ A. Henden, G. Walker\end{tabular}} \\ \hline
\multicolumn{1}{|l|}{\begin{tabular}[c]{@{}l@{}}La Palma, \\ Spain\end{tabular}} &
  \multicolumn{1}{l|}{\begin{tabular}[c]{@{}l@{}}28.762516,\\ -17.8792,\\ 2387.63\end{tabular}} &
  \multicolumn{1}{l|}{\begin{tabular}[c]{@{}l@{}}Liverpool,\\ 2.0,\\ -\end{tabular}} &
  \multicolumn{1}{l|}{\begin{tabular}[c]{@{}l@{}}PC+NTP,\\ RISE,\\ -\end{tabular}} &
  \multicolumn{1}{l|}{\begin{tabular}[c]{@{}l@{}}1.183,\\ 1.2355,\\ -\end{tabular}} &
  \multicolumn{1}{l|}{\begin{tabular}[c]{@{}l@{}}N. Morales, \\ PI. R. Duffard\end{tabular}}\\ \hline
\end{tabular}
}
\caption{Observational circumstances of all observatories that obtained a positive detection in the three stellar occultations by the Huya system.}
\label{tab:observational_circumstances_positive}
\end{table}

\begin{table}[!htb]
\resizebox{\linewidth}{!}{
\begin{tabular}{|llllll|}
\hline
\multicolumn{1}{|c|}{\textbf{\begin{tabular}[c]{@{}c@{}}Observatory,\\ nearest city,\\ country\end{tabular}}} &
  \multicolumn{1}{c|}{\textbf{\begin{tabular}[c]{@{}c@{}}Latitude ($^\circ$),\\ longitude ($^\circ$),\\ height (m)\end{tabular}}} &
  \multicolumn{1}{c|}{\textbf{\begin{tabular}[c]{@{}c@{}}Telescope,\\ aperture (m),\\ filter\end{tabular}}} &
  \multicolumn{1}{c|}{\textbf{\begin{tabular}[c]{@{}c@{}}Time source,\\ instrument,\\ written time\end{tabular}}} &
  \multicolumn{1}{c|}{\textbf{\begin{tabular}[c]{@{}c@{}}Exposure (s),\\ cycle (s)\\ result\end{tabular}}} &
  \multicolumn{1}{c|}{\textbf{Observers}} \\ \hline
\multicolumn{6}{|c|}{2021-03-28}\\ \hline
\multicolumn{1}{|l|}{\begin{tabular}[c]{@{}l@{}}Wise,\\ Mitzpe Ramon,\\ Israel\end{tabular}} &
  \multicolumn{1}{l|}{\begin{tabular}[c]{@{}l@{}}30.5958333,\\ 34.76333,\\ 857.0\end{tabular}} &
  \multicolumn{1}{l|}{\begin{tabular}[c]{@{}l@{}}C28 prime focus,\\ 0.71,\\ Luminance\end{tabular}} &
  \multicolumn{1}{l|}{\begin{tabular}[c]{@{}l@{}}PC +  NTP,\\ ProLine PL16803,\\ Start of Exposure\end{tabular}} &
  \multicolumn{1}{l|}{\begin{tabular}[c]{@{}l@{}}10.0,\\ 12.1,\\ Negative \end{tabular}} &
  \multicolumn{1}{l|}{\begin{tabular}[c]{@{}l@{}}S. Kaspi \end{tabular}} \\ \hline
\multicolumn{6}{|c|}{2023-02-17}\\ \hline
\multicolumn{1}{|l|}{\begin{tabular}[c]{@{}l@{}}Garner State Park\\ Texas,\\ Unites States of America\end{tabular}}&
  \multicolumn{1}{|l|}{\begin{tabular}[c]{@{}l@{}}29.5969,\\ -99.7328,\\ 443.0\end{tabular}} &
  \multicolumn{1}{|l|}{\begin{tabular}[c]{@{}l@{}}Newtonian,\\ 0.315,\\ None\end{tabular}} &
  \multicolumn{1}{l|}{\begin{tabular}[c]{@{}l@{}}IOTA-VTI,\\ WAT-910HX-RC,\\ Middle of Exposure\end{tabular}} &
  \multicolumn{1}{l|}{\begin{tabular}[c]{@{}l@{}}1.068,\\ 1.068,\\ Negative \end{tabular}} &
  \multicolumn{1}{l|}{\begin{tabular}[c]{@{}l@{}} S. Messner \end{tabular}} \\ \hline
\multicolumn{1}{|l|}{\begin{tabular}[c]{@{}l@{}}Nederland\\ Colorado,\\ Unites States of America\end{tabular}} &
  \multicolumn{1}{l|}{\begin{tabular}[c]{@{}l@{}}39.98720968649829,\\ -105.4455682399913,\\ 2492.62\end{tabular}} &
  \multicolumn{1}{l|}{\begin{tabular}[c]{@{}l@{}}Skywatcher,\\ 0.20,\\ None\end{tabular}} &
  \multicolumn{1}{l|}{\begin{tabular}[c]{@{}l@{}}GPS,\\ QHY174M-GPS,\\ Start of Exposure\end{tabular}} &
  \multicolumn{1}{l|}{\begin{tabular}[c]{@{}l@{}}4.0,\\ 4.32,\\ Negative \end{tabular}} &
  \multicolumn{1}{l|}{\begin{tabular}[c]{@{}l@{}} M. Skrutskie,\\Anne J. Verbiscer\\ \end{tabular}} \\ \hline
  \multicolumn{6}{|c|}{2023-06-24}\\ \hline
\multicolumn{1}{|l|}{\begin{tabular}[c]{@{}l@{}}IOTA Scorpii,\\ Italy\end{tabular}} &
  \multicolumn{1}{l|}{\begin{tabular}[c]{@{}l@{}}44.12703306956611,\\ 9.856022392325654,\\ 52.0\end{tabular}} &
  \multicolumn{1}{l|}{\begin{tabular}[c]{@{}l@{}}GSO 16,\\ 0.406,\\ Clear\end{tabular}} &
  \multicolumn{1}{l|}{\begin{tabular}[c]{@{}l@{}}PC+GPS,\\ STXL6303E,\\ Start of Exposure\end{tabular}} &
  \multicolumn{1}{l|}{\begin{tabular}[c]{@{}l@{}}15.0,\\ 18.68,\\ Negative\end{tabular}} &
  \multicolumn{1}{c|}{G. Scarfi} \\ \hline
\multicolumn{1}{|l|}{\begin{tabular}[c]{@{}l@{}}Belesta,\\ France\end{tabular}} &
  \multicolumn{1}{l|}{\begin{tabular}[c]{@{}l@{}}43.445408,\\ 1.8175,\\ 247.397\end{tabular}} &
  \multicolumn{1}{l|}{\begin{tabular}[c]{@{}l@{}}Newtonian,\\ 0.82,\\ Gaia Clear (G)\end{tabular}} &
  \multicolumn{1}{l|}{\begin{tabular}[c]{@{}l@{}}PC+NTP,\\ C3-PRO-61000 - CMOS,\\ Start of Exposure\end{tabular}} &
  \multicolumn{1}{l|}{\begin{tabular}[c]{@{}l@{}}1.0,\\ 3.0,\\ Negative\end{tabular}} &
  \multicolumn{1}{l|}{\begin{tabular}[c]{@{}l@{}}P. Martinez,\\ P. André\end{tabular}} \\ \hline
\multicolumn{1}{|l|}{\begin{tabular}[c]{@{}l@{}}Guirguillano,\\ Navarra,\\ Spain\end{tabular}} &
  \multicolumn{1}{l|}{\begin{tabular}[c]{@{}l@{}}42.712053,\\ -1.865,\\ 594.0\end{tabular}} &
  \multicolumn{1}{l|}{\begin{tabular}[c]{@{}l@{}}Sultán,\\ 0.31,\\ Empty\end{tabular}} &
  \multicolumn{1}{l|}{\begin{tabular}[c]{@{}l@{}}GPS,\\ QHY174M,\\ Start of Exposure\end{tabular}} &
  \multicolumn{1}{l|}{\begin{tabular}[c]{@{}l@{}}10.0\\ 10.008\\ Negative \end{tabular}} &
  \multicolumn{1}{l|}{\begin{tabular}[c]{@{}l@{}}J. Prat\\ P. Martorell\end{tabular}} \\ \hline
\multicolumn{1}{|l|}{\begin{tabular}[c]{@{}l@{}}Otivar,\\ Andalucia,\\ Spain\end{tabular}} &
  \multicolumn{1}{l|}{\begin{tabular}[c]{@{}l@{}}36.81611111880951,\\ -3.6802975274933303,\\ 314.365478515625\end{tabular}} &
  \multicolumn{1}{l|}{\begin{tabular}[c]{@{}l@{}}ASA 12,\\ 0.30,\\ R\end{tabular}} &
  \multicolumn{1}{l|}{\begin{tabular}[c]{@{}l@{}}PC+NTP,\\ ZWO ASI1600MM,\\ Middle of Exposure\end{tabular}} &
  \multicolumn{1}{l|}{\begin{tabular}[c]{@{}l@{}}5.0,\\ 5.45,\\ Negative\end{tabular}} &
  \multicolumn{1}{l|}{\begin{tabular}[c]{@{}l@{}}A. Popowicz \\ \&\\ SUTO Team\end{tabular}}\\ \hline
\multicolumn{1}{|l|}{\begin{tabular}[c]{@{}l@{}}Črni Vrh,\\ Slovenia\end{tabular}} &
  \multicolumn{1}{l|}{\begin{tabular}[c]{@{}l@{}}45.94585244301794,\\ 14.071284495949174,\\ 713.0279541015625\end{tabular}} &
  \multicolumn{1}{l|}{\begin{tabular}[c]{@{}l@{}}Cichocki - Astrograph,\\ 0.60,\\ W\end{tabular}} &
  \multicolumn{1}{l|}{\begin{tabular}[c]{@{}l@{}}PC+NTP,\\ ZWO ASI6200MM,\\ Start of Exposure\end{tabular}} &
  \multicolumn{1}{l|}{\begin{tabular}[c]{@{}l@{}}5.0,\\ 5.9,\\ Negative\end{tabular}} &
  \multicolumn{1}{l|}{H. Mikuz}\\ \hline
\multicolumn{1}{|l|}{\begin{tabular}[c]{@{}l@{}}Starhopper, \\ Covasna,\\ Romania\end{tabular}} &
  \multicolumn{1}{l|}{\begin{tabular}[c]{@{}l@{}}45.865556,\\ 25.768889,\\ 588.0\end{tabular}} &
  \multicolumn{1}{l|}{\begin{tabular}[c]{@{}l@{}}Meade 16 LX200,\\ 0.406,\\ None\end{tabular}} &
  \multicolumn{1}{l|}{\begin{tabular}[c]{@{}l@{}}PC+NTP,\\ Canon 6D,\\ Start of Exposure\end{tabular}} &
  \multicolumn{1}{l|}{\begin{tabular}[c]{@{}l@{}}6.0,\\ 7.0,\\ Inconclusive\end{tabular}} &
  \multicolumn{1}{l|}{F. Ursache} \\ \hline
\multicolumn{1}{|l|}{\begin{tabular}[c]{@{}l@{}}Albox,\\ Spain\end{tabular}} &
  \multicolumn{1}{l|}{\begin{tabular}[c]{@{}l@{}}37.405564,\\ -2.1518,\\ 491.0\end{tabular}} &
  \multicolumn{1}{l|}{\begin{tabular}[c]{@{}l@{}}Meade16,\\ 0.406,\\ Clear\end{tabular}} &
  \multicolumn{1}{l|}{\begin{tabular}[c]{@{}l@{}}PC+GPS,\\ Atik314L+,\\ Start of Exposure\end{tabular}} &
  \multicolumn{1}{l|}{\begin{tabular}[c]{@{}l@{}}7.0,\\ 8.03,\\ Negative\end{tabular}} &
  \multicolumn{1}{l|}{J. L. Maestre} \\ \hline
\multicolumn{1}{|l|}{\begin{tabular}[c]{@{}l@{}}Estelia,\\ Asturias,\\ Spain\end{tabular}} &
  \multicolumn{1}{l|}{\begin{tabular}[c]{@{}l@{}}43.20358286,\\ -5.4449518,\\ 630.0\end{tabular}} &
  \multicolumn{1}{l|}{\begin{tabular}[c]{@{}l@{}}Ritchey–Chrétien,\\ 0.30,\\ No filter\end{tabular}} &
  \multicolumn{1}{l|}{\begin{tabular}[c]{@{}l@{}}PC+GPS,\\ QHY268M,\\ Start of Exposure\end{tabular}} &
  \multicolumn{1}{l|}{\begin{tabular}[c]{@{}l@{}}15.0,\\ 15.008,\\ Negative\end{tabular}} &
  \multicolumn{1}{l|}{\begin{tabular}[c]{@{}l@{}}E. Fernández, \\ N. Graciá\end{tabular}} \\ \hline
\multicolumn{1}{|l|}{\begin{tabular}[c]{@{}l@{}}BOOTES-1,\\ Huelva,\\ Spain\end{tabular}} &
  \multicolumn{1}{l|}{\begin{tabular}[c]{@{}l@{}}37.10408250638086,\\ -6.734117424827392,\\ 61.0\end{tabular}} &
  \multicolumn{1}{l|}{\begin{tabular}[c]{@{}l@{}}BOOTES-1b,\\ 0.30,\\ Clear\end{tabular}} &
  \multicolumn{1}{l|}{\begin{tabular}[c]{@{}l@{}}PC+NTP,\\ DV897\_BV\_BOOTES-1b,\\ Start of Exposure\end{tabular}} &
  \multicolumn{1}{l|}{\begin{tabular}[c]{@{}l@{}}20.0,\\ 21.7,\\ Negative\end{tabular}} &
  \multicolumn{1}{l|}{\begin{tabular}[c]{@{}l@{}}I. Perez-Garcia, \\ PI: A. Castro-Tirado\end{tabular}}\\ \hline
\multicolumn{1}{|l|}{\begin{tabular}[c]{@{}l@{}}BOOTES-2, \\ Malaga, \\ Spain\end{tabular}} &
  \multicolumn{1}{l|}{\begin{tabular}[c]{@{}l@{}}36.759241,\\ -4.04097,\\ 70.0\end{tabular}} &
  \multicolumn{1}{l|}{\begin{tabular}[c]{@{}l@{}}BOOTES-2,\\ 0.60,\\ Clear\end{tabular}} &
  \multicolumn{1}{l|}{\begin{tabular}[c]{@{}l@{}}PC+NTP,\\ Andor Ixon EMCCD DU8201\_BV,\\ Start of Exposure\end{tabular}} &
  \multicolumn{1}{l|}{\begin{tabular}[c]{@{}l@{}}30.0,\\ 30.007,\\ Negative\end{tabular}} &
  \multicolumn{1}{l|}{\begin{tabular}[c]{@{}l@{}}E. J. Fernandez-Garcia, \\ PI: A. Castro-Tirado\end{tabular}}\\ \hline
\multicolumn{1}{|l|}{\begin{tabular}[c]{@{}l@{}}Sant Esteve Sesrovires, \\ Catalonia,\\ Spain\end{tabular}} &
  \multicolumn{1}{l|}{\begin{tabular}[c]{@{}l@{}}41.49361,\\ 1.8725,\\ 180.0\end{tabular}} &
  \multicolumn{1}{l|}{\begin{tabular}[c]{@{}l@{}}Newtonian,\\ 0.40,\\ Empty\end{tabular}} &
  \multicolumn{1}{l|}{\begin{tabular}[c]{@{}l@{}}IOTA-VTI,\\ WATEC-910HX/RC,\\ Start of Exposure\end{tabular}} &
  \multicolumn{1}{l|}{\begin{tabular}[c]{@{}l@{}}5.08\\ 5.08\\ Inconclusive\end{tabular}} &
  \multicolumn{1}{l|}{C. Schnabel} \\ \hline
\multicolumn{1}{|l|}{\begin{tabular}[c]{@{}l@{}}Calar Alto, \\ Spain\end{tabular}} &
  \multicolumn{1}{l|}{\begin{tabular}[c]{@{}l@{}}37.22361,\\ -2.5461,\\ 2168.0\end{tabular}} &
  \multicolumn{1}{l|}{\begin{tabular}[c]{@{}l@{}}SCT,\\ 1.23,\\ Empty\end{tabular}} &
  \multicolumn{1}{l|}{\begin{tabular}[c]{@{}l@{}}PC+NTP,\\ ASI 461,\\ Start of Exposure\end{tabular}} &
  \multicolumn{1}{l|}{\begin{tabular}[c]{@{}l@{}}0.8,\\ 0.864,\\ Negative\end{tabular}} &
  \multicolumn{1}{l|}{N. Morales} \\ \hline
\multicolumn{1}{|l|}{\begin{tabular}[c]{@{}l@{}}Sierra Nevada, \\ Granada,\\ Spain\end{tabular}} &
  \multicolumn{1}{l|}{\begin{tabular}[c]{@{}l@{}}37.064136,\\ -3.3847,\\ 2930.527\end{tabular}} &
  \multicolumn{1}{l|}{\begin{tabular}[c]{@{}l@{}}T90,\\ 0.90,\\ Empty\end{tabular}} &
  \multicolumn{1}{l|}{\begin{tabular}[c]{@{}l@{}}PC+NTP,\\ QHY600M-L,\\ Start of Exposure\end{tabular}} &
  \multicolumn{1}{l|}{\begin{tabular}[c]{@{}l@{}}2.0,\\ 2.0,\\ Negative\end{tabular}} &
  \multicolumn{1}{l|}{\begin{tabular}[c]{@{}l@{}}F. J. Aceituno,\\ PI: P. Santos-Sanz\end{tabular}} \\ \hline
\multicolumn{1}{|l|}{\begin{tabular}[c]{@{}l@{}}**San Marcello Pistoiese, \\ Italy\end{tabular}} &
  \multicolumn{1}{l|}{\begin{tabular}[c]{@{}l@{}}44.063036,\\ 10.8042,\\ 965.411\end{tabular}} &
  \multicolumn{1}{l|}{\begin{tabular}[c]{@{}l@{}}Marcon,\\ 0.60,\\ Unfilter\end{tabular}} &
  \multicolumn{1}{l|}{\begin{tabular}[c]{@{}l@{}}PC+NTP,\\ Apogee,\\ Start of Exposure\end{tabular}} &
  \multicolumn{1}{l|}{\begin{tabular}[c]{@{}l@{}}6.0,\\ 7.0,\\ Negative\end{tabular}} &
  \multicolumn{1}{l|}{\begin{tabular}[c]{@{}l@{}}P. Bacci, \\ M. Maestripieri, \\ M. D. Grazia\end{tabular}}\\ \hline
\multicolumn{1}{|l|}{\begin{tabular}[c]{@{}l@{}}**Asiago Astrophysical,\\ Asiago, \\ Italy\end{tabular}} &
  \multicolumn{1}{l|}{\begin{tabular}[c]{@{}l@{}}45.849444,\\ 11.568824,\\ 1370.0\end{tabular}} &
  \multicolumn{1}{l|}{\begin{tabular}[c]{@{}l@{}}Schmidt,\\ 0.91,\\ Clear\end{tabular}} &
  \multicolumn{1}{l|}{\begin{tabular}[c]{@{}l@{}}PC+NTP,\\ Moravian,\\ Start of Exposure\end{tabular}} &
  \multicolumn{1}{l|}{\begin{tabular}[c]{@{}l@{}}7.0,\\ 12.0,\\ Negative\end{tabular}} &
  \multicolumn{1}{l|}{\begin{tabular}[c]{@{}l@{}}D. Nardiello, \\ V. Nascimbeni\end{tabular}} \\ \hline
\multicolumn{1}{|l|}{\begin{tabular}[c]{@{}l@{}}PixelSkies, \\ Granada, \\ Spain\end{tabular}} &
  \multicolumn{1}{l|}{\begin{tabular}[c]{@{}l@{}}37.74002,\\ -2.64395,\\ 850.0\end{tabular}} &
  \multicolumn{1}{l|}{\begin{tabular}[c]{@{}l@{}}EdgeHD 11,\\ 0.279,\\ UV/IR\end{tabular}} &
  \multicolumn{1}{l|}{\begin{tabular}[c]{@{}l@{}}PC+NTP,\\ ASI2400MC Pro,\\ Start of Exposure\end{tabular}} &
  \multicolumn{1}{l|}{\begin{tabular}[c]{@{}l@{}}-\\ -\\ Technical failure\end{tabular}} &
  \multicolumn{1}{l|}{V. Pelenjow} \\ \hline
\multicolumn{1}{|l|}{\begin{tabular}[c]{@{}l@{}}Sarriguren, \\ Navarra,\\ Spain\end{tabular}} &
  \multicolumn{1}{l|}{\begin{tabular}[c]{@{}l@{}}42.80833800489953,\\ -1.5895521640777588,\\ 462.5870361328125\end{tabular}} &
  \multicolumn{1}{l|}{\begin{tabular}[c]{@{}l@{}}Meade LX200-ACF,\\ 0.203,\\ IR-UV\end{tabular}} &
  \multicolumn{1}{l|}{\begin{tabular}[c]{@{}l@{}}Other,\\ Asi 1600MM,\\ End of Exposure\end{tabular}} &
  \multicolumn{1}{l|}{\begin{tabular}[c]{@{}l@{}}-\\ -\\ Technical failure\end{tabular}} &
  \multicolumn{1}{l|}{M. A. A. Amat} \\ \hline
\multicolumn{1}{|l|}{\begin{tabular}[c]{@{}l@{}}Monte Agliale, \\ Garfagnana,\\ Italy\end{tabular}} &
  \multicolumn{1}{l|}{\begin{tabular}[c]{@{}l@{}}43.99528,\\ 10.51486,\\ 760.0\end{tabular}} &
  \multicolumn{1}{l|}{\begin{tabular}[c]{@{}l@{}}Lotti,\\ 0.50,\\ Empty\end{tabular}} &
  \multicolumn{1}{l|}{\begin{tabular}[c]{@{}l@{}}PC+NTP,\\ SBIG ST9,\\ Start of Exposure\end{tabular}} &
  \multicolumn{1}{l|}{\begin{tabular}[c]{@{}l@{}}-\\ -\\ Overcast\end{tabular}} &
  \multicolumn{1}{l|}{F. Ciabattari} \\ \hline
\multicolumn{1}{|l|}{\begin{tabular}[c]{@{}l@{}}Forcarei,\\ Spain\end{tabular}} &
  \multicolumn{1}{l|}{\begin{tabular}[c]{@{}l@{}}42.610591,\\ -8.37088,\\ 670.063\end{tabular}} &
  \multicolumn{1}{l|}{\begin{tabular}[c]{@{}l@{}}RCOS,\\ 0.50,\\ Clear\end{tabular}} &
  \multicolumn{1}{l|}{\begin{tabular}[c]{@{}l@{}}Other,\\ WATEC 910HX RC,\\ Start of Exposure\end{tabular}} &
  \multicolumn{1}{l|}{\begin{tabular}[c]{@{}l@{}}Technical\\ issues\\ -\end{tabular}}&
  \multicolumn{1}{l|}{H. González-Rodriguez} \\ \hline
\end{tabular}
}
\caption{Observational circumstances for the other sites that attempted or acquired data during the stellar occultation campaigns. The ** symbol means that original images were not provided, only the light curve made by the observer.}
\label{tab:huya_observational_circumstances1}
\end{table}

\newpage
\section{Reanalysis of 2019 data}
\label{sec:appendix_2019}
Here, we present the \textit{2019 Restrict} limb solution as mentioned in the text. As the image sets are not publicly available, to perform the limb fitting, we used the \textsc{sora} v0.3.1 Python library and instants with their uncertainties as published by \cite{SantosSanz2022}. Without the original information, it is impossible to distinguish between positive chords' bad times and topography in the object's profile. Therefore, in this work, we choose to use the data as they are; e.g., we did not consider topography in Huya and neither applied offsets to the positive chords. For instance, the LC1 and LC2 come from telescopes in the same observatory, but they do not agree with each other. Therefore, such misalignments may be the reason for the large $\chi^2_{pdf}$ presented in Table \ref{tab:fitted parameters}.

\begin{figure}[!htb]
    \centering
    \includegraphics[width=\linewidth]{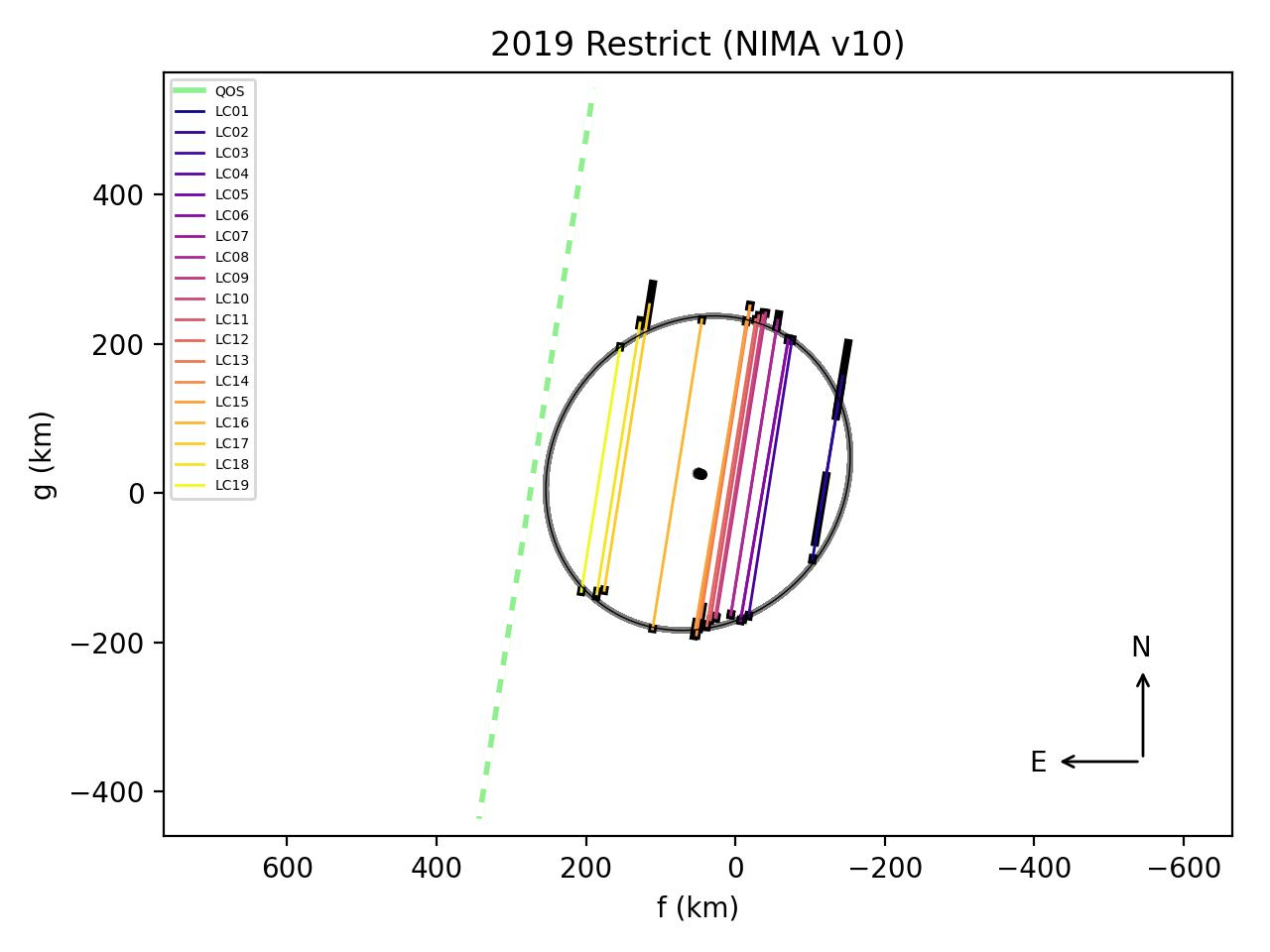}
    \caption{Huya's limb as observed during the stellar occultation in 2019-03-28 \citep{SantosSanz2022}. The dashed green line shows the negative data set observed from the QOS Observatory in Ukraine. The solid, colorful lines represent the observed positive light curves (LC), following the same definition as in the original publication. Black segments represent the published uncertainties in the star dis- and re-appearance instants. The black ellipse represents the \textit{2019 Restrict} limb solution presented in Table \ref{tab:fitted parameters}. The gray area shows the solution's $3\sigma$ uncertainty.}
    \label{fig:2019_limb_restrict}
\end{figure}

\newpage
\section{System density determination}
\label{sec:appendix_density}
The system mass was derived from the mutual orbit presented in this work, so the system density can be obtained assuming a Maclaurin shape for Huya with axes being $a=b=a'$. This way we obtained the true semimajor and semiminor axes $a=b=218.05 \pm 0.11$ km with uncertainties coming from the ellipse fitted to the 2019 data set. The object's true oblateness is calculated as follows \citep{Braga-Ribas2013}, 
\begin{equation}
    \epsilon = 1- \frac{\sqrt{(R_{eq}/a')^4-cos^2(\theta)}}{sin(\theta)} = 0.14,
    \label{eq:epsilon_true}
\end{equation}
where $\theta$ is the polar axis aspect angle and $R_{eq} = a' \sqrt{1-\epsilon'}$. In this work, we assumed an equatorial orbit for the satellite, so we have $\theta = 60^\circ \pm 3.5^\circ$. The uncertainty comes from the partial derivatives, as follows
\begin{equation}
\delta \epsilon = \sqrt{\left(\frac{\partial \epsilon}{\partial R_{eq}} \delta R_{eq}\right)^2+\left(\frac{\partial \epsilon}{\partial a} \delta a \right)^2+\left(\frac{\partial \epsilon}{\partial \theta} \delta \theta\right)^2}=0.011.
\end{equation}

Once the true oblateness is obtained, the true polar axis with uncertainty can be calculated by 
\begin{equation}
    c=a(1-\epsilon)=187.5\mathrm{\ km}
\end{equation}
and
\begin{equation}
    \delta c = \sqrt{((1-\epsilon)\delta a)^2+((-a)\delta\epsilon)^2} = 2.4 \mathrm{\ km.}
\end{equation}

Huya volume can then be obtained from 
\begin{equation}
    V_{Huya} = \frac{4}{3}\pi abc ,
    \label{eq:huya_vol}
\end{equation}
where the $a$, $b$, and $c$ are Huya's true semi-major axes obtained before.

The satellite has a minimum spherical diameter of D = 165 km from the most accurate single-chord detection and a maximum diameter of D = 243 km from the published values obtained from thermal measurements. This provides a minimum and a maximum volume for a spherical body as follows
\begin{equation}
    V_{Sat} = \frac{4}{3}\pi R^3, 
    \label{eq:huya_vol}
\end{equation}
where $R$ is the minimum or maximum radius of the spherical satellite. Finally, the density for the Huya system are $\rho_{1}$ = 1073 $\pm$ 66 kg~m$^{-3}$.

\newpage
\section{Orbit Fitting Outputs}
\label{sec:appendixe}

\begin{figure}[!htb]
    \centering
    \includegraphics[width=0.99\linewidth]{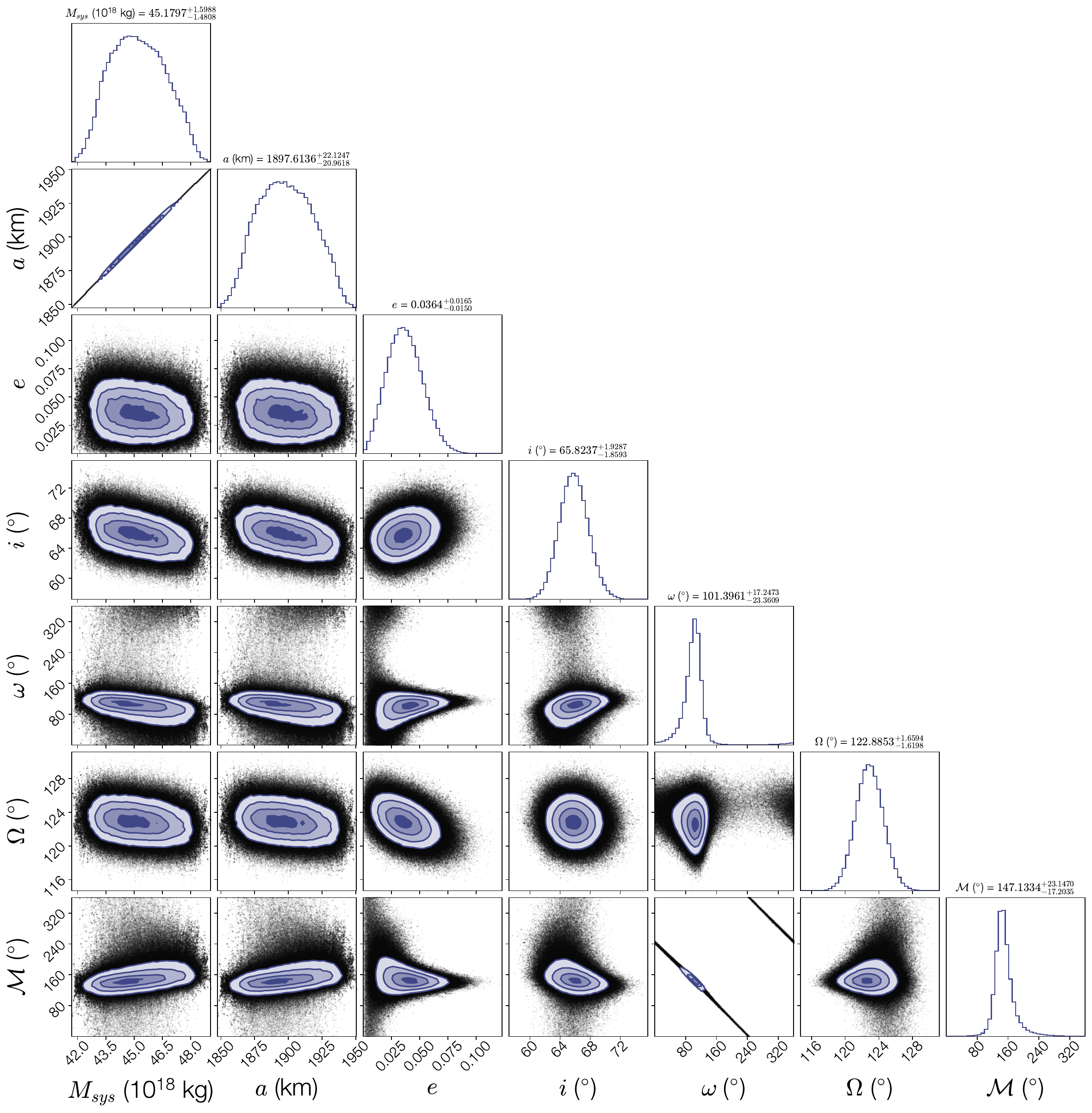}
    \caption{A corner plot of the MCMC chains from the satellite orbit fit. The 2-dimensional contour plots show the joint posterior distributions for each pair of parameters, and the histograms at the top of each column show the marginal posteriors of each parameter. Black points show individual samples from the MCMC chains.
    }
    \label{fig:corner}
\end{figure}

Here, we show a corner plot output from the orbit fitting process (Figure \ref{fig:corner}). Joint posterior distributions are shown as 2-dimensional contour plots, and marginal posteriors are shown as histograms at the top of each column. 

\newpage
\bibliography{sample631}{}
\bibliographystyle{aasjournal}

\end{document}